\documentclass[reprint,amsmath,amssymb,aps,floatfix,onecolumn,superscriptaddress]{revtex4-2}

\usepackage[dvipsnames]{xcolor}

\usepackage{hyperref}
\hypersetup{linktocpage,
	pdftoolbar=true,        
	pdfmenubar=true,        
	pdffitwindow=false,     
	pdfstartview={FitH},    
	pdftitle={title},    
	pdfauthor={Gerard Valentí-Rojas},     
	pdfsubject={Subject},   
	pdfcreator={Gerard Valentí-Rojas},   
	pdfproducer={Gerard Valentí-Rojas}, 
	pdfnewwindow=true,      
	colorlinks=true,       
	linkcolor=teal,          
	citecolor=OrangeRed,        
	filecolor=Green,      
	urlcolor=teal           
}

\usepackage[bitstream-charter]{mathdesign}
\DeclareSymbolFont{usualmathcal}{OMS}{cmsy}{m}{n}
\DeclareSymbolFontAlphabet{\mathcal}{usualmathcal}
\DeclareMathAlphabet\mathbfcal{OMS}{cmsy}{b}{n}

\usepackage[inline]{enumitem}
\usepackage{graphicx}
\usepackage{ifthen}
\usepackage{amsmath}
\usepackage{amsthm}
\usepackage{pifont}
\usepackage{array}
\usepackage{siunitx}
\usepackage{xstring}
\usepackage{multirow}
\usepackage{datetime}
\usepackage{etoolbox}

\usepackage{slashed}

\usepackage{bm}
\usepackage{braket}

\usepackage{booktabs}

\usepackage{subcaption}

\usepackage{cleveref}

\newtheorem{thm}{Claim}

\usepackage{framed}

\definecolor{shadecolor}{rgb}{0.95,0.95,0.95}

\newenvironment{asidebox}[1][\hsize]
{%
    \MakeFramed{\hsize#1\advance\hsize-\width\FrameRestore}%
}
{\endMakeFramed}

\newenvironment{example}[1][\hsize]
{%
    \MakeFramed{\hsize#1\advance\hsize-\width\FrameRestore}%
}
{\endMakeFramed}

\usepackage{graphicx}
\usepackage{dcolumn}
\usepackage{url}

\begin{document}
\title{The Composite Particle Duality: A New Class of Topological Quantum Matter}
\author{Gerard Valent\'i-Rojas}
\email{gerard.valenti-i-rojas@universite-paris-saclay.fr}

\affiliation{Naquidis Center, Institut d'Optique Graduate School, 91127, Palaiseau, France}
\affiliation{Institut de Mathématiques de Bordeaux, UMR 5251, Université de Bordeaux, France}
\affiliation{SUPA, Institute of Photonics and Quantum Sciences,
Heriot-Watt University, Edinburgh, EH14 4AS, United Kingdom}

\author{Joel Priestley}
\affiliation{SUPA, Institute of Photonics and Quantum Sciences,
Heriot-Watt University, Edinburgh, EH14 4AS, United Kingdom}

\author{Patrik \"Ohberg}
\affiliation{SUPA, Institute of Photonics and Quantum Sciences,
Heriot-Watt University, Edinburgh, EH14 4AS, United Kingdom}

\date{\today}
\begin{abstract}
The composite particle duality extends the notions of both flux attachment and statistical transmutation in spacetime dimensions beyond 2+1D. It constitutes an exact correspondence that can be understood either as a theoretical framework or as a dynamical physical mechanism. The immediate implication of the duality is that an interacting quantum system in arbitrary dimensions can experience a modification of its statistical properties if coupled to a certain gauge field. In other words, commutation relations of quantum fields can be effectively modified by a dynamical physical process. For instance, an originally bosonic quantum fluid in $d$ spatial dimensions can feature composite fermionic (or anyonic) excitations when coupled to a statistical gauge field. In 3+1D the mechanism of flux attachment induces a dynamical formation of dyons as higher-dimensional analogues of Laughlin quasiparticles. In 1+1D there is lack of flux attachment but a remnant in the form of a statistical gauge field can be explicitly constructed. We also introduce a family of interacting quantum many-body systems that undergo statistical transmutation as indicated by the duality. This opens the door to a new realm of topological phases across dimensions both in lattice and continuum systems.
\end{abstract}
\maketitle



\section{Changing the Identity of Quantum Matter}
\label{sec:change}

It was Leinaas and Myrheim \cite{leinaas1977theory} who challenged the symmetrisation postulate approach in Quantum Mechanics and identified important loopholes in the allowed quantum statistics of identical particles. In proving that there was no need for a symmetrisation postulate in order to obtain Bose and Fermi statistics for \textit{point} particles in spatial dimensions $d\ge 3$, they found that intermediate statistics were possible in $d<3$. In addition, for spatial dimension greater than one, they argued that a flat vector potential would give rise to a gauge transformation that could alter, what they called, the ``many-body nature'' of the system. Nowadays, we call this transformation \textit{statistical transmutation} \cite{polyakov1988fermi}. The work of Leinaas and Myrheim was, in essence, an existence proof allowing possibilities that had been overlooked since the very birth of Quantum Mechanics. Their groundbreaking results, however, were framed in a seemingly academic setup, with no reference to a physical mechanism, nor concrete examples of such a transmutation. It was not until the advent of Chern-Simons theory \cite{schonfeld1981mass,deser1982topomass} and the discovery of the fractional quantum Hall effect (FQHE) \cite{tsui198fqhe,laughlin1983anomalous} that this was given a physical meaning in the form of an emergent phenomenon in a strongly-correlated system \cite{girvin1987off,zhang1989effective,read1989order}. The low-energy excitations of a solid could become \textit{anyons} despite the fundamental constituents being electrons, i.e. fermions. Parallel, and superficially independent developments with regards to unusual behaviours of quantum statistics, appeared also in other contexts. Theoretical work in  high energy physics suggested that spin-statistics properties of dyons could be altered \cite{hasenfratz1975fermion,goldhaber1976chargemono} in the presence of a $\theta-$vacuum. On the other hand, works on the Luttinger liquid and the development of bosonization techniques seemed to point out that there is a direct non-local mapping between bosonic and fermionic fields \cite{lieb1965exact,luther1974,haldane1981luttinger,mandelstam1975soliton,coleman1975sine}, a Bose-Fermi correspondence. All the previous phenomena happen in different spacetime dimensions and look disconnected from one another, should it not be for the sole fact that there is statistical transmutation at play.\\

At the time of writing, it appears to be early days for a general theory and universal understanding of fractionalisation, although some aspects of it, like charge fragmentation, are fairly well understood \cite{jackiw1976spin,su1979solitons,goldston1981fractional,su1981fractionally,backinblack1985,jackiw2007graphene,chamon2008electron,chamon2008irrational,moessner2021topological,fradkin2013}. For instance, there seems to be consensus in that Dirac fermions coupled to topological defects often fractionalise fermion number. This consensus is not present in the case of statistical transmutation, which is not discussed much beyond Chern-Simons theory in two spatial dimensions. We now recognise statistical transmutation as a manifestation of fractionalisation of statistics. Moreover, there appears to be enough evidence to infer that the effective statistics of matter is not an immutable property, but can actually be altered and maybe even controlled \cite{valenti20synthetic,valenti2024dual}. It is then natural to ask whether \textit{a physical mechanism, encoded in a mathematical mapping, can drive effective Bose-Fermi correspondences across different dimensions}. \\

\begin{figure}[h]
\centering
    \includegraphics[width=0.9\textwidth]{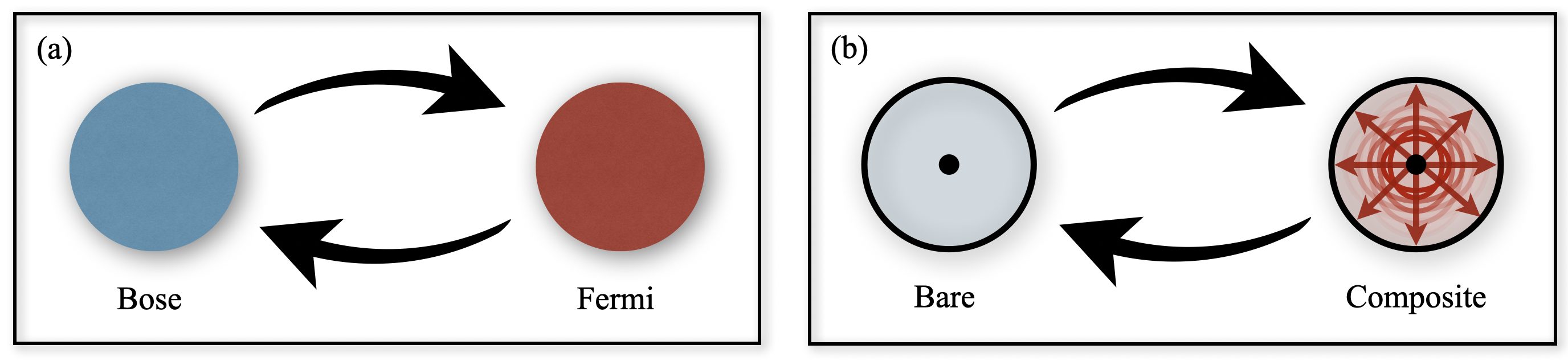}
    \caption{(a) Bose-Fermi correspondences and (b) Bare-Composite dualities as introduced in this work. We might see some of these correspondences as the consequence of a statistical gauge field dressing a bare entity and effectively inducing statistical transmutation.}
 \label{fig:barecompo}
\end{figure}

In this work we attempt to give an answer to the previous question. We argue that there exists at least one physical mechanism, namely the \textit{composite particle duality} introduced in Ref. \cite{valenti2023composite}, that naturally identifies a notion of a statistical gauge field which alters the identity of matter in spatial dimensions other than two, a case for which statistical transmutation via a Chern-Simons mechanism is well under control. A stepping stone of our approach is the success of flux attachment \cite{wilckez1982flux,wilczek1982fraction} in explaining and providing an intuition for statistical transmutation in the FQHE and beyond. Both these concepts, together with the associated composite particle picture \cite{fradkin2013,ezawa2013quantum,jain2007composite}, we argue, are notions that transcend the context in which they were originally conceived. We explicitly generalise both flux attachment and statistical transmutation separately only to bring them together in the broader framework of an exact physical correspondence in which matter can change its quantum identity by being appropriately dressed by statistical gauge fields in any spatial dimension.\\

Based on the proposed correspondence, we introduce a family of interacting quantum many-body models in several dimensions that accept a straightforward local composite dual description implying statistical transmutation. These systems can be formulated both as lattice models or continuum non-relativistic field theories. A subset of the models introduced corresponds to well-known instances of topological quantum matter, others are new. They are all theories of matter coupled to statistical gauge fields in the appropriate dimension, for which exotic phenomenology is expected. \\

The outline of this work is as follows. We start by motivating the problem and reviewing some necessary concepts and notation in Section \ref{sec:prelim}. We introduce the composite particle duality in Section \ref{sec:comp_part_du} after separately generalising flux attachment, statistical transmutation and the Jordan-Wigner transformation. We examine in detail the cases of one, two and three spatial dimensions. In light of this, in Section \ref{sec:sitelink} we are able to introduce some families of models for topological quantum matter with a particularly simple composite dual description both in continuum and on the lattice. We conclude this work in Section \ref{sec:summary} by emphasising some of the universal aspects of our proposed framework and highlighting the potential for future work and extensions, especially for relativistic systems.\\

For the sake of clarity, we have included box environments which summarise or highlight key concepts. Spacetime coordinates and dimensionality are denoted as $x = (t,\mathbf{x})$ and $D=d+1$ respectively throughout this work. 

\section{Preliminaries}
\label{sec:prelim}

Before we start with the discussion of our work, it is important to motivate why it is sensible to even consider a general mechanism for statistical transmutation in the first place. The exotic statistics of quantum particles allowed in the plane can be traced back to the configuration space $\mathcal{H}_{n}$ for $n$ point-like particles, which is multiply-connected in $2d$. In particular, its fundamental or first homotopy group corresponds to the \textit{braid} group $\pi_{1}(\mathcal{H}^{d=2}_{n}) = \mathcal{B}_{n}\,$, which gives mathematical soundness to the existence of anyons as point-like objects. This should be contrasted with a similar exercise performed for point-like particles in $d\ge 3$, for which the first homotopy group is the permutation group $\pi_{1}(\mathcal{H}^{d\ge 3}_{n}) = S_{n}$ and whose one-dimensional representations are bosons and fermions. One can also show that point-like particles in three spatial dimensions satisfy the conventional spin-statistics relation \cite{luders1958}. This is, by now, textbook material on the subject \cite{simon2023topological,pachos2012introduction}. However, there are some important subtleties and remarks that are worth making explicit.
\begin{asidebox}
(i) First, statistical transmutation has traditionally been intertwined with the presence of anyons but, more fundamentally, it concerns the possibility of altering the statistics of particles. A Bose-Fermi (or Fermi-Bose) transmutation \cite{goldhaber1976chargemono,hasenfratz1975fermion,jackiw1976spin,balachandran1987quantum,forte1991boson} is allowed in all dimensions regardless of the possible existence or not of exotic statistics. Thus, we might wonder what kind of process could alter the quantum identity of matter across dimensions.\\

(ii) Secondly, statistics of particles are purely quantum-mechanical properties as they are linked to the indistinguishability of identical objects at small scales. By extension, statistical transmutation understood as a change in this purely quantum identity is also a quantum effect. This is supported by the evidence that in the well-known case of two spatial dimensions, the key physics at play are those of the Aharonov-Bohm effect acting as a topological Berry phase. This is also not something fundamentally tied to dimensionality.\\

(iii) Finally, there is, by now, substantial theoretical evidence supporting the possibility of anyons in higher dimension not as point-like, but as extended  objects \cite{freedman2011projec,ye2015vortexline,wang2019loops,aneziris1991novel,baez2007exotic,bergeron1995canonical}. The locality postulate assumed in the previous topological arguments is violated and the arguments no longer hold. In a condensed matter context, one could envision string or loop-like quasiparticle excitations arising in $d=3$ with exotic statistics under mutual exchange. The mechanism giving rise to such objects might be physically related to remarks (i) and (ii), but it is logically independent.   
\end{asidebox}

Our approach will rely on basic notions of gauge fields minimally coupling to matter, so we briefly revisit some concepts that should be nothing new to the reader, but a gentle introduction to the new material. 

\subsection{To Remove or Not To Remove... Gauge Fields}
A gauge transformation connects locally different gauge connections $A\,(x)$ and $A'(x)$ giving rise to the same curvature or field strength $F(x)$. In practice, this amounts to adding a total derivative to the gauge field to find a gauge-transformed configuration without altering the Physics. Certain configurations are equivalent --- or can be deformed --- to the identity. For an Abelian 1-form gauge field in $D=d+1$ spacetime dimensions, a conventional gauge transformation is given by
\begin{equation}
    A_{\mu}(x)\;\;\; \longrightarrow\;\;\; A_{\mu}'(x) = \mathcal{U}(x)\, A_{\mu}(x)\,\mathcal{U}^{-1}(x) - \frac{i}{g} [\partial_{\mu}\,\mathcal{U}(x)]\; \mathcal{U}^{-1}(x) =  A_{\mu}(x) + \frac{1}{g}\,\partial_{\mu}\xi(x)\;,
\end{equation}
where $\mu=0,\dots,d$ indicates spacetime components. We find the field strength $F_{\mu \nu}\,(x) = \partial_{\mu} \,A_{\nu}(x) - \partial_{\nu} \,A_{\mu}(x)$ to be gauge-invariant. Furthermore, matter couples to gauge fields via a gauge-covariant derivative, which transforms like
\begin{equation}
    D_{\mu} \Psi (x) \;\;\; \longrightarrow\;\;\;  D'_{\mu} \Psi'(x) = \mathcal{U}(x)\,\Big[\partial_{\mu} -i g A_{\mu}(x)\Big]\,\Psi(x)\;,
\end{equation}
where the matter field transforms like $\Psi'(x) = \mathcal{U}(x)\, \Psi(x)$, while for the conjugate field $\bar{\Psi}'(x) = \bar{\Psi}(x) \,\mathcal{U}^{\dagger}(x)$. Now, since the gauge transformation is given by a unitary matrix $\mathcal{U}^{-1}(x) = \mathcal{U}^{\dagger}(x)$, kinetic terms in the associated Lagrangians and Hamiltonians are invariant to gauge transformations. If the gauge field can be expressed in the form \footnote{Here $\tilde{\mathcal{U}}$ is simply used to indicate that it is a particular case of the general case $\mathcal{U}$. This is merely a stylistic point.}
\begin{equation}
\tilde{A}_{\mu}(x) = -i\,[\partial_{\mu}\,\mathcal{\tilde{U}}(x)]\;\mathcal{\tilde{U}}^{-1}(x) = \partial_{\mu} \phi (x)
\end{equation}
where $\phi(x)$ is some well-defined scalar field, $\tilde{A}_{\mu}$ is referred to as a topologically trivial \textit{pure gauge} configuration. The associated curvature is $F_{\mu\nu} = 0$, so it does not give rise to field strengths --- e.g. electromagnetic fields --- and the connection is said to be flat. In addition, it can be gauge transformed to $A_{\mu}'(x)= 0$. That is, it can be \textit{removed} or \textit{trivialised} via a gauge transformation.\\

We imagine now that the field $\phi(x)$ is not well-defined everywhere. For instance, it can have a singularity at some spacetime point $x = x_{0}$, which we signal by $\phi(x) \equiv \phi_{s}(x; x_{0})$. Then, the above mechanism still holds everywhere except for precisely at $x_{0}$ as the gauge field and transformation become ill-defined. Here, we say that the gauge field is locally, but not globally, a pure gauge, or that the connection is non-trivially flat. More generally, the consequence of these ill-defined points, lines, surfaces, etc. is that they become seeds for topological invariants, i.e. countable objects to wrap around. When these defects cannot be naïvely removed \footnote{It is important to notice here that the crucial elements are both the dimensionality of the defect and that of the configuration space.}, they provide \textit{obstructions} to the removal of the gauge field, we then say that the gauge field becomes ``topological''. Fully removing these gauge fields requires a \textit{large gauge transformation} parametrised by $\mathcal{W}(x)$, which does not connect physically equivalent states, but configurations with different homotopic properties yielding the same observable consequences. This means that matter fields transform like $\Psi'(x) = \mathcal{W}(x)\, \Psi(x)\;,$ but $\mathcal{W}(x) \equiv \mathcal{W}(x;x_{0}) = \exp{\big[i\phi_{s}(x;x_{0})\big]}$ now carries defects, as they are present in the scalar field $\phi(x)$. This has important implications as these features are transferred to the transformed matter field $\Psi'(x) \equiv \Psi'(x,x_{0})$, implying that such a gauge transformation has the ability to add (or remove) defects. Hence, even though we believe we remove topologically non-trivial pure gauge fields from a system, this comes at a cost of altering the matter field, so it is not truly \textit{removed} but \textit{hidden}. This redefinition of the matter field can also be thought of as a \textit{dressing} process. This type of transformation is recurrently found in literature, not as a gauge transformation, but as a mere identity, redefinition, or mapping of matter fields. In this context, $\mathcal{W}(x)$ might be referred to as a \textit{disorder field} \cite{kadanoffceva1971}. For quantised theories, it is promoted to a \textit{disorder operator}. Depending on the specific type of defect it contains, it is typically known as a soliton, kink, vortex or monopole operator, amongst others. If $\Psi(x)$ is an order parameter or operator, we say that the mapping has an order-disorder operator structure \cite{hasenfratz1979puzzling,fradkin2017disorder,marino2017quantum}. This is a recurrent structure in Bose-Fermi dualities \cite{fradkin2013}. The reason is relatively simple and rarely spelled out in literature. If we consider the equal-time (anti)commutation relations $[\bullet,\bullet]_{\,\mp}$ of initially bosonic (fermionic) matter fields, we can apply said transformation to obtain
\begin{flalign}
   & \big[\hat{\Psi}'(x),\hat{\Psi}'(x')\big]_{\,\mp} = \big[\hat{\mathcal{W}}(x)\,\hat{\Psi}(x),\hat{\mathcal{W}}(x')\,\hat{\Psi}(x')\big]_{\,\mp} = 
   \hat{\mathcal{W}}(x)\,\big[\hat{\Psi}(x),\hat{\mathcal{W}}(x')\big]_{\,-}\,\hat{\Psi}(x') \\ &+ \hat{\mathcal{W}}(x) \hat{\mathcal{W}}(x') \,\big[\hat{\Psi}(x),\hat{\Psi}(x')\big]_{\,-} +  \big[\hat{\mathcal{W}}(x),\hat{\mathcal{W}}(x')\big]_{\,-}\,\hat{\Psi}(x') \hat{\Psi}(x) + \hat{\mathcal{W}}(x')\, \big[\hat{\mathcal{W}}(x),\hat{\Psi}(x')\big]_{\,\mp} \hat{\Psi}(x)
\end{flalign}
and 
\begin{flalign}
  &  \big[\hat{\Psi}'(x),\hat{\bar{\Psi}}'(x')\big]_{\,\mp} = \big[\hat{\mathcal{W}}(x)\,\hat{\Psi}(x),\hat{\bar{\Psi}}(x')\,\hat{\mathcal{W}}^{\dagger} (x')\big]_{\,\mp} = \hat{\mathcal{W}}(x)\,\big[\hat{\Psi}(x),\hat{\bar{\Psi}}(x')\big]_{\,-}\,\hat{\mathcal{W}}^{\dagger}(x') \\ &+ \hat{\mathcal{W}}(x) \hat{\bar{\Psi}}(x') \,\big[\hat{\Psi}(x),\hat{\mathcal{W}}^{\dagger}(x')\big]_{\,-} +  \big[\hat{\mathcal{W}}(x),\hat{\bar{\Psi}}(x')\big]_{\,-}\, \hat{\mathcal{W}}^{\dagger}(x') \hat{\Psi}(x) + \hat{\bar{\Psi}}(x')\, \big[\hat{\mathcal{W}}(x),\hat{\mathcal{W}}^{\dagger}(x')\big]_{\,\mp} \hat{\Psi}(x)\;.
\end{flalign}
Therefore, we observe that the statistics of the original fields are transformed and depend on the statistics of the disorder operator. We also notice that these reduce to the usual relations when $\hat{\mathcal{W}}(x) =\hat{\mathbb{I}}$. If the disorder operators trivially commute with themselves and with the matter fields, the original commutation relations are preserved up to re-scaling of the matter fields. It is now no mystery how to get a fermionic theory out of a bosonic one (or \textit{vice versa}), a suitable disorder operator dressing the appropriate matter fields will do the job. We emphasise that such a transformation can be defined without any reference to it being a large gauge transformation, but merely a local canonical transformation. We also draw attention to the fact that we have neither assumed an explicit dimensionality for our system, nor even a particular type of model.\\

The previous argument on gauge transformation will allow us to design a physical mechanism out of the mathematical mapping that we just witnessed. The disorder operator will become a large gauge transformation that will introduce or remove a minimally-coupled non-trivial pure gauge field.

\section{The Composite Particle Duality}
\label{sec:comp_part_du}

The previous discussion illustrates how some exact equivalences can effectively modify the identity of charged particles, i.e. transmute their exchange statistics. We now argue that these transformations appear naturally from the coupling to emergent gauge fields with topological content in a quantum system of many particles. Naïvely, there is nothing preventing such a notion from being general, regardless of dimensionality, the original statistics of the system, and whether it is found on the lattice or in continuum. It is worth noting that the specific statistical nature of the composite fields will depend on dimensionality and locality, but the ability to go from bare to composite will not.\\

We argue that an interacting many-body system can self-generate effective statistical gauge fields in arbitrary dimensions. These gauge fields not only couple to matter but also bind to it and, in doing so, they \textit{dress it} so that it appears to change its intrinsic properties. This should be no different than an interaction-driven topological many-body Berry phase effect \cite{valenti20synthetic}. In the following, we provide evidence portraying how such a physical mechanism can work. For that, it becomes necessary to disentangle the notions of \textit{statistical transmutation} and \textit{flux attachment}, which will be generalised independently to be then connected and interpreted in a broader framework. The former is the ability for fields to dynamically change their statistics, while the latter is the mechanism by which matter nucleates gauge flux.

\subsection{Generalised Statistical Transmutation by a Statistical Gauge Potential}
The notion of statistical transmutation was introduced by Polyakov \cite{polyakov1988fermi} as the connection between Bose and Fermi matter fields mediated by a Chern-Simons term. Here, we define statistical transmutation more loosely as the ability to dynamically change the properties under exchange or, equivalently, modify the (anti)commutation relations of quantum matter fields. A well-known example appears in the context of FQHE, where it is stated as a theorem \cite{zhang1995chern} for the particular case of a Chern-Simons vector potential in two spatial dimensions. In the following, we show an equivalence between non-interacting theories with different statistical properties under exchange of particles in arbitrary dimensions. We might think of this as a \textit{generalised statistical transmutation} process. This is formulated in first-quantised formalism to be later translated to second-quantised language and extended to interacting systems.
\begin{asidebox}
\noindent 
Let us consider two distinct non-relativistic non-interacting quantum many-body systems, namely B (Bare) and C (Composite). While theory B satisfies bosonic/fermionic phase factors under exchange of particle labels, theory C satisfies generalised exchange relations specified by a statistical factor $\gamma$.
\begin{thm}
There exists an exact equivalence between theories B and C  sharing the same eigenvalues. Therefore, B and C give rise to the same physics but can have different statistical properties. The statistical factor $\gamma$ might or might not be reduced to a sign ($\pm$) upon additional constraints on the problem.  
\end{thm}  
\end{asidebox}

In layman's terms, a system of locally interacting charged particles in the presence of a gauge field is equivalent to a different system of charged particles where no gauge field is present, but whose many-body wavefunction has changed along with some of its properties. We work with $\hbar = c = q = 1$ and consider theory B in $d$ spatial dimensions as 
\begin{equation}\label{eq:theory_b}
	\mathcal{H}_{\text{\,B}}\,\Psi\,(\mathbf{x}_{1},\dots,\mathbf{x}_{N}) = E\,\Psi\,(\mathbf{x}_{1},\dots,\mathbf{x}_{N})\;\;\;\;\;\; \;\;\;\;\;\; \text{with} \;\;\;\;\;\; \mathcal{H}_{\text{\,B}} = \sum_{i=1}^{N} \frac{\big[\mathbf{p}\,(\mathbf{x}_{i}) - \bm{a}\,(\mathbf{x}_{i})\big]^{\,2}}{2m}\;,
\end{equation}
where $\bm{a}(\mathbf{x}_{i}) \equiv \bm{a}(\mathbf{x}_{i})\,[\mathbf{x}_{1},\dots,\mathbf{x}_{N}] $ is an Abelian gauge potential, whose form is to be specified, such that a particle located at point $\mathbf{x}_{i} = (x^{1}_{i},\dots,x^{d}_{i})$ experiences the presence of the rest of particles. Consequently, its value can depend on the positions of all $N$ particles in the system, and therefore is a many-body gauge field. The wavefunction satisfies
\begin{equation}
	\Psi\,(\mathbf{x}_{1},\dots,\mathbf{x}_{i},\dots,\mathbf{x}_{j},\dots,\mathbf{x}_{N}) = \pm\, \Psi\,(\mathbf{x}_{1},\dots,\mathbf{x}_{j},\dots,\mathbf{x}_{i},\dots,\mathbf{x}_{N})
\end{equation}
under the exchange of two particles, where the sign depends on the bosonic/symmetric ($+$) or fermionic/antisymmetric ($-$) nature of the wavefunction. We claim the existence of another theory C, equivalent to the previous one, of the form
\begin{equation}
	\tilde{\mathcal{H}}_{\text{\,C}}\,\Psi_{c}\,(\mathbf{x}_{1},\dots,\mathbf{x}_{N}) = E\,\Psi_{c}\,(\mathbf{x}_{1},\dots,\mathbf{x}_{N})\;\;\;\;\;\; \;\;\;\;\;\; \text{with} \;\;\;\;\;\; \tilde{\mathcal{H}}_{\text{\,C}} = \sum_{i=1}^{N} \frac{\tilde{\mathbf{p}}\,(\mathbf{x}_{i}) ^{\,2}}{2m}
\end{equation}
where the wavefunction satisfies
\begin{equation}\label{eq:general_exchange}
	\Psi_{c}\,(\mathbf{x}_{1},\dots,\mathbf{x}_{i},\dots,\mathbf{x}_{j},\dots,\mathbf{x}_{N}) = \pm\, e^{\,i\gamma}\, \Psi_{c}\,(\mathbf{x}_{1},\dots,\mathbf{x}_{j},\dots,\mathbf{x}_{i},\dots,\mathbf{x}_{N}) \;,
\end{equation}
and where $\gamma$ is a \textit{statistical factor} (or \textit{angle}). We shall prove that it is possible to obtain $\tilde{\mathcal{H}}_{C}$ from $\mathcal{H}_{B}$, and that such a transformation alters the exchange properties of the many-body wavefunction, giving rise to the $\gamma$ factor. This is known as \textit{statistical transmutation}. Let us then introduce a \textit{duality (gauge) transformation} of the form 
\begin{equation}\label{eq:transform_first}
	\Psi\,(\mathbf{x}_{1},\dots,\mathbf{x}_{N}) = \mathcal{W}\,(\mathbf{x}_{1},\dots,\mathbf{x}_{N})\, \Psi_{c}\,(\mathbf{x}_{1},\dots,\mathbf{x}_{N}) \;\;\;\;\;\; \;\;\;\;\;\; \text{with} \;\;\;\;\;\; \mathcal{W}\,(\mathbf{x}_{1},\dots,\mathbf{x}_{N})  = e^{\,i\alpha \,\Phi\,(\mathbf{x}_{1}, \dots,\mathbf{x}_{N})}
\end{equation}
where we keep $\Phi$ unspecified and $\alpha$ is a real-valued parameter. Substituting Eq. \eqref{eq:transform_first} in Eq. \eqref{eq:theory_b} and using $\mathbf{p}\,(\mathbf{x}_{i}) = -i\,\bm{\nabla}_{\mathbf{x}_{i}}$, we are left with
\begin{equation}
	\mathcal{H}_{\text{\,B}}\,\Psi\,(\mathbf{x}_{1},\dots,\mathbf{x}_{N}) =	\frac{1}{2m}\, \Big[ \sum_{i=1}^{N} \big( -i\,\bm{\nabla}_{\mathbf{x}_{i}} - \bm{a}\,(\mathbf{x}_{i}) \big)^{2}\Big] \cdot \Big[ \mathcal{W}\,(\mathbf{x}_{1},\dots,\mathbf{x}_{N})\, \Psi_{c}\,(\mathbf{x}_{1},\dots,\mathbf{x}_{N})\Big]\;.
\end{equation}
Using the notation $  \bm{a}_{\mathbf{x}_{i}} \equiv \bm{a} \,(\mathbf{x}_{i})$, $\mathcal{W}$, $\Phi$ and $\Psi_{c}$, as well as the property $\bm{\nabla}_{\mathbf{x}_{i}} \mathcal{W} = i\alpha\, \mathcal{W}\,\bm{\nabla}_{\mathbf{x}_{i}} \Phi\,$, we obtain
\begin{equation}
	\frac{\mathcal{W}}{2m}\, \bigg\{ \sum_{i=1}^{N} \Big[- \bm{\nabla}_{\mathbf{x}_{i}}^{2} \Psi_{c} + 2i\,\Big(\,\bm{a}_{\mathbf{x}_{i}} -\alpha \bm{\nabla}_{\mathbf{x}_{i}} \Phi  \Big) \,\bm{\nabla}_{\mathbf{x}_{i}} \Psi_{c} + \Big( (\bm{a}_{\mathbf{x}_{i}} -\alpha\,\bm{\nabla}_{\mathbf{x}_{i}}\Phi)^{2} - i\alpha \bm{\nabla}_{\mathbf{x}_{i}}^{2} \Phi + i\,\bm{\nabla}_{\mathbf{x}_{i}} \bm{a}_{\mathbf{x}_{i}}  \Big) \,\Psi_{c} \Big]\bigg\}\;.
\end{equation}
We  observe that the second ($\sim \nabla_{\mathbf{x}_{i}} \Psi_{c}$ ) and third ($\sim \Psi_{c}$ ) terms in brackets have to vanish in order to recover
\begin{equation}
	\frac{\mathcal{W}}{2m} \sum_{i=1}^{N} \Big[- \bm{\nabla}_{\mathbf{x}_{i}}^{2} \Psi_{c} \Big] =  \mathcal{W}\,\tilde{\mathcal{H}}_{C}\Psi_{c} = E\,\Psi \;,
\end{equation}
and thus
\begin{equation}
\tilde{\mathcal{H}}_{\text{\,C}}\,\Psi_{c}\,(\mathbf{x}_{1},\dots,\mathbf{x}_{N}) = E\,\Psi_{c}\,(\mathbf{x}_{1},\dots,\mathbf{x}_{N}) \;.
\end{equation}
Therefore, solving the above constraints as differential equations for $\bm{a}$ and $\Phi$ allows this mapping to be true. However, we observe a cancellation of both terms for the constraint
\begin{equation}\label{eq:pure_gauge}
	\bm{a}_{\mathbf{x}_{i}} \,(\mathbf{x}_{1},\dots,\mathbf{x}_{N})  = \alpha \,\bm{\nabla}_{\mathbf{x}_{i}} \Phi\,(\mathbf{x}_{1},\dots,\mathbf{x}_{N}) \;.
\end{equation}
This tells us that the vector potential has the form of a \textit{pure gauge} and proves that this mapping works regardless of the value of $\Phi$. Recall both the vector potential and the transformation have been kept as free parameters. This concludes the first part of the proof, which neither contains approximations, \textit{nor does it depend on fixing an a priori dimensionality of the system}. It is solely the statement that pure gauges can be removed by an appropriate gauge transformation which, in this case, is given by $\mathcal{W}$.\\

It remains then to be shown that Eq. \eqref{eq:general_exchange} is satisfied upon substitution of the duality transformation. In essence, that in the process of removing the gauge potential, the new many-body wave function acquires exotic exchange properties. For that, we will define $\mathcal{W}^{-1} = \exp\,(-i\alpha\,\Phi^{*})$ and use properties $\mathcal{W}^{-1} \mathcal{W} = 1$, $\mathcal{W}^{-1} = \mathcal{W^{*}}$ and $\Phi = \Phi^{*}$, to find
\begin{equation}
	\Psi_{c}\,(\mathbf{x}_{1},\dots,\mathbf{x}_{i},\dots,\mathbf{x}_{j},\dots,\mathbf{x}_{N}) = \pm\, e^{\,- i\alpha \,\Phi\,(\mathbf{x}_{1},\dots,\mathbf{x}_{i},\dots,\mathbf{x}_{j},\dots,\mathbf{x}_{N})}e^{\,i\alpha \,\Phi\,(\mathbf{x}_{1},\dots,\mathbf{x}_{j},\dots,\mathbf{x}_{i},\dots,\mathbf{x}_{N})}\, \Psi_{c}\,(\mathbf{x}_{1},\dots,\mathbf{x}_{j},\dots,\mathbf{x}_{i},\dots,\mathbf{x}_{N}) \;.
\end{equation}
Let us assume for now that the duality transformation can be decomposed in a sum of two-body functions in the form $\Phi \,(\mathbf{x}_{1},\dots,\mathbf{x}_{N})  = \sum_{ m < l} \phi\,(\mathbf{x}_{m},\mathbf{x}_{l}) \,$. This allows writing
\begin{equation}
	\mathcal{W}\,(\mathbf{x}_{1},\dots,\mathbf{x}_{N}) = \prod_{m<l} e^{i\alpha \,\phi\,(\mathbf{x}_{m},\mathbf{x}_{l})} \equiv \prod_{m<l} w\,(\mathbf{x}_{m},\mathbf{x}_{l})\;.
\end{equation}
Note that if both $m$ \textit{and} $l$ are different from $i$ \textit{or} $j$, the phase factors cancel out and we are left with
\begin{equation}\label{eq:exchange}
	\Psi_{c}\,(\mathbf{x}_{1},\dots,\mathbf{x}_{i},\dots,\mathbf{x}_{j},\dots,\mathbf{x}_{N}) = \pm\, e^{-\,i\alpha \,\big(\sum_{m<l}^{'} \phi_{ml} \,-\, \sum_{p<q}^{'} \phi_{p q}\big)} \, \Psi_{c}\,(\mathbf{x}_{1},\dots,\mathbf{x}_{j},\dots,\mathbf{x}_{i},\dots,\mathbf{x}_{N}) \;,
\end{equation}
where we use the notation $\phi_{ml} \equiv \phi\,(\mathbf{x}_{m},\mathbf{x}_{l})$. The primes ($'$) denote that sums are taken over $l$ and $q$ indices respectively that are equal to $i$ or $j$. Also, for every $\phi_{ml}$ there is a $\phi_{pq} = \phi_{lm}\,$. Furthermore, we see that the decomposition in terms of two-body functions is a sufficient but not a necessary condition, i.e. it is chosen for simplicity. A similar logic would apply if we considered a more general three-body decomposition like $\Phi \,(\mathbf{x}_{1},\dots,\mathbf{x}_{N})  = \sum_{m < l < k} \phi\,(\mathbf{x}_{m},\mathbf{x}_{l},\mathbf{x}_{k}) $ or $n$-body decomposition for $n<N$, so that $\Phi \,(\mathbf{x}_{1},\dots,\mathbf{x}_{N})  = \sum_{m < l < k < \dots < n} \phi\,(\mathbf{x}_{m},\mathbf{x}_{l},\mathbf{x}_{k},\dots,\mathbf{x}_{n}) $ , but then the treatment rapidly becomes tedious.\\

At this point we can already identify theories B and C as equivalent provided the pure gauge constraint [Eq. \eqref{eq:pure_gauge}] is satisfied, and where we identify the statistical factor as
\begin{equation}
    \gamma = - \alpha \,\Big(\sum_{m<l}' \phi_{ml} \,-\, \sum_{p<q}' \phi_{p q}\Big)
\end{equation}
with $\phi_{ml}$ being real-valued functions yet to be specified. We can now consider further properties of $\phi_{ml}$ --- or, equivalently $w_{ml}$ --- in order to find explicit values for $\gamma$. Natural choices are:
\begin{itemize}
    \item \textbf{Symmetric exchange.} We consider even parity upon exchange of labels $\phi_{ml} = \phi_{lm}\,$. This choice yields $\gamma_{s} = 0$, and it corresponds to a trivial removal of a pure gauge. Statistics are not altered.

    \item \textbf{Antisymmetric exchange.} We consider odd parity upon exchange of labels $\phi_{ml} = - \phi_{lm}\,$. Here, $\gamma_{a} \equiv \gamma_{a}(\mathbf{x}_{m},\mathbf{x}_{l}) = - 2\alpha \sum_{m<l}' \phi_{ml}\,$. There is explicit dependence on the pure gauge function.

    \item \textbf{Linear exchange.} We assume a property under exchange of the form $\phi_{ml} = \beta_{1} + \beta_{2}\, \phi_{lm}$ where $\beta_{1},\beta_{2} \in [0,2\pi)$. The particular case of $\beta_{2} = +1$ yields $\gamma_{l} = \alpha \beta_{1}\in [0,2\pi)$, meaning that boson-fermion (fermion-boson) transmutation is achieved for the particular case $\alpha \beta_{1} = \pi + 2\pi k$ with $k\in \mathbb{Z}$. More generally, $\alpha \beta_{1}$ define an anyonic-type exchange, with the caveat that additional constraints to these values can come from further considerations of physical consistency, e.g. a conventional spin-statistics relation for point-like objects in $d=3$.
\end{itemize}
\vspace{0.3cm}
In all the cases above, the choice is not unique, as we are only specifying the exchange properties of $\phi_{ml}$, but not their explicit functional form. This implies that the route to statistical transmutation from bare to composite is also non-unique. In other words, there are several vector potentials that give rise to the same statistical factor. This can be understood as gauge freedom and such a family of vector potentials forms a statistical gauge orbit.\\

\begin{figure}[h]
\centering
    \includegraphics[width=0.8\textwidth]{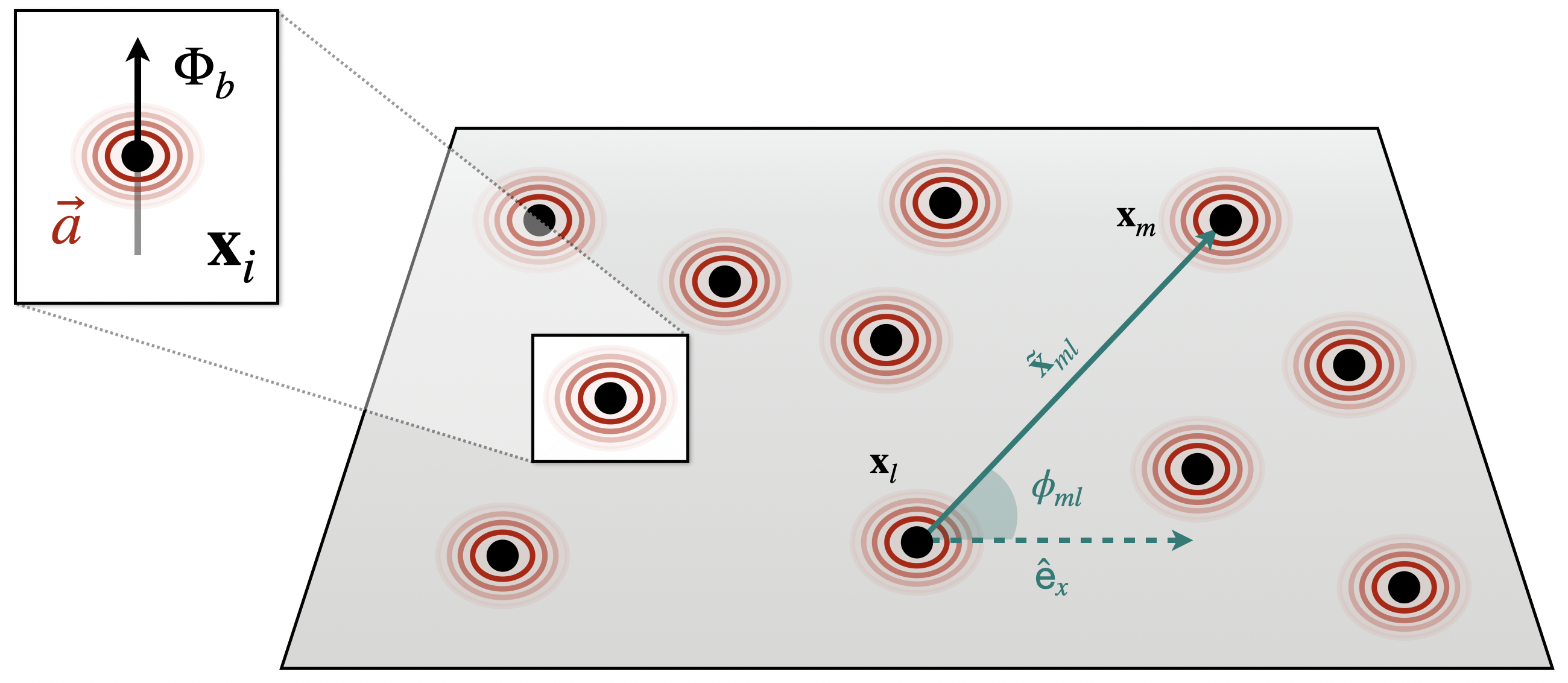}
    \caption{Conventional flux attachment on the plane. A gas of charge-flux-tube composites interacting via many-body Aharonov-Bohm effect. Point-like electric charges have each locally captured a magnetic flux $\Phi_{b}$ and sit at the centre of a vortex vector potential $\vec{a}\,(\mathbf{x}_{i})$ that decays as $\sim 1/r\,$, depicted by the concentric circles radially fading away from the charges. The statistical phase factor is obtained upon summing over all the pairwise contributions $\phi_{ml}$ corresponding to mutual Aharonov-Bohm phases.}
 \label{fig:manyaharon}
\end{figure}

As a closing remark of this section, we have shown that an Abelian vector potential can be removed in a non-interacting system in $d$ spatial dimensions at the expense of changing the two-particle exchange relations of the many-body wavefunction, thus connecting two superficially different theories. It is sufficient for the total vector potential to be
\begin{equation}
   \bm{a}\,(\mathbf{x}) = \sum_{i=1}^{N} \bm{a}\,(\mathbf{x}_{i}) = \alpha\,\sum_{i=1}^{N}\bm{\nabla}_{\mathbf{x}_{i}} \,\Big[ \sum_{a<b=1}^{N} \phi\,(\mathbf{x}_{a},\mathbf{x}_{b})\Big]\;,
\end{equation}
to perform statistical transmutation. It is therefore called a \textit{statistical gauge potential}. The statistical properties of the composite theory manifestly depend on the exchange properties of the many-body pure gauge function $\phi_{ab}\,$. The interested reader is referred to Appendix \ref{sec:snippets} for a discussion of practical cases. Later on we will see how this picture is preserved also for interacting systems.

\subsection{Generalised Flux Attachment by a Topological Gauge Field} \label{sec:genFluxsec}
We define flux attachment in 2+1D as a \textit{local} constraint or Gauss's law between magnetic field and charge (or number) density $\bm{\nabla}\times \bm{a}\,(t,\mathbf{x}) = \gamma\,n\,(t,\mathbf{x})$. We can also define it \textit{globally} over a closed surface $\Sigma$ as a proportionality relation between magnetic flux and electric charge (or number of particles) contained in that region as $\Phi_{b}^{\Sigma} = \gamma\, Q_{\Sigma}$. It is evident that flux attachment is to be expected in 2n+1D for $n\in \mathbb{Z}^{+}$, from the mere existence of a Chern-Simons term in odd spacetime dimensions. However, the global formulation of the constraint suggests that it might be universal across dimensions $d\ge 2$, since magnetic flux and charges can be defined in all generality. By this token, having bare electric charges that act as a source of magnetic flux is something that transcends dimensions. We can also wonder about the $d=1$ case. In essence, what is left of flux attachment when a magnetic field cannot even be defined? Can there be statistical transmutation?\\

The success of Chern-Simons theory has taught us that when such a topological field theory minimally couples to matter --- i.e. a conserved Noether current $\mathcal{J}^{\mu}(x)$ ---, flux attachment emerges in the associated Euler-Lagrange equations. Solving them for the associated Chern-Simons gauge field gives us its functional form, in a given gauge, as a function of matter. This gauge field acts as a statistical gauge field begetting the transmutation of statistics. Hence, flux attachment not only generates electric-magnetic composites through Gauss's law, but also fixes the form of the statistical gauge field. In other words, it provides a physical mechanism that naturally manufactures the unknown gauge field that we need to insert in the previous section and that, up until now, remained unknown. Our point of departure to extend flux attachment to arbitrary dimensions is precisely the Abelian Chern-Simons theory. In covariant form, the flux attachment constraint is a topological current of the form $\mathcal{J}^{\mu}(t,\mathbf{x}) \propto \epsilon^{\,\mu \nu \lambda}\,\partial_{\nu} a_{\lambda}(t,\mathbf{x})$. It is not difficult to formulate a $p$-form extension of this topological current law, namely
\begin{equation}\label{eq:gen_flat}
    \star \mathcal{J}^{\,(p)}(x) = \mathcal{F}^{\,(D-p)}(x)= \text{d}\mathcal{A}^{(D-p-1)}(x) \;.
\end{equation}
This is a \textit{Hodge duality} mapping $p-$forms in $\text{D}$ dimensions to $(\text{D}-p)-$forms through the Hodge star operator $(\star)\,$. We can check that, in $D=3$ and for a $p=1$ gauge form, this precisely coincides with the Chern-Simons result. Expression \eqref{eq:gen_flat} can be found rather systematically by assuming the existence of a conserved Noether current satisfying $\star \,\text{d}\mathcal{J} = 0$. Poincaré's lemma guarantees the existence of a form $\mathcal{A}$ that solves this conservation equation yielding \eqref{eq:gen_flat} as both the solution and the definition of a topological current. Furthermore it ensures the existence of solutions upon transformations of the form
\begin{equation}\label{eq:high_gt}
    \mathcal{A}^{(p)}(x)\rightarrow\mathcal{A}^{(p)}(x) + \text{d}\xi^{(p-1)}(x)
\end{equation}
which we now interpret as proof of gauge redundancy, and Eq. \eqref{eq:high_gt} as a higher form of antisymmetric gauge transformation. We can write the $(\text{D}-2)-$form field as $\mathcal{A} = \text{d} \alpha$ at the expense of violating the corresponding Bianchi identity $\text{d}(\text{d}\alpha) \neq 0$ due to non-trivial topology of $\alpha\,$. This motivates the introduction of $\mathcal{A}\,(x)$ as a $p$-form topological gauge connection or, more loosely, a \textit{topological gauge field}. We claim that Eq. \eqref{eq:gen_flat} constitutes the $D$-dimensional extension of conventional flux attachment. Now, we might insist that the Noether current remains a $D$-dimensional Lorentz vector, which constrains $\mathcal{A}$ to be a $(D-2)$ gauge-dependent form. In covariant field theoretical language, and introducing a coupling constant $\gamma$, we might rewrite the previous expressions as follows.
\begin{asidebox}
\begin{thm}
There exists a generalised $D$-dimensional expression for flux attachment of the local form  
\begin{equation}\label{eq:generalised_flux}
    \mathcal{J}^{\mu}(t,\mathbf{x}) = \gamma\, \epsilon^{\,\mu \nu \lambda \alpha \beta \dots}\,\partial_{\nu} \mathcal{A}_{\lambda \alpha \beta \dots}(t,\mathbf{x})\;,
\end{equation}
where $\mathcal{A}\,$ is a rank-$(\text{D}-2)$ antisymmetric $\text{U}(1)$ gauge field that locally transforms like 
\begin{equation}
\mathcal{A}_{\mu\nu\lambda\alpha \beta \dots}(t,\mathbf{x}) \;\longrightarrow \; \mathcal{A}_{\mu\nu\lambda\alpha \beta \dots}(t,\mathbf{x}) + \partial_{\,[\mu}\,\xi_{\,\nu\lambda\alpha \beta \dots]}(t,\mathbf{x})\;\;,
\end{equation}
and $\gamma$ is a real constant whose value may or may not be constrained through Dirac quantisation.
\end{thm}
\end{asidebox}
\noindent We observe that the central object is not the Chern-Simons gauge field, but in general, an antisymmetric tensorial topological gauge field. The rank $p$ of the topological gauge field is tied to the dimensionality $D$ of the problem. This means that:
\begin{itemize}
    \item In $D=1+1$ the topological gauge field is a $0$-form, i.e. a scalar $\mathcal{A}^{(0)} = \phi (t,\mathbf{x})$.
    \item In $D=2+1$ it is a $1$-form, i.e. a vector --- a.k.a. a Chern-Simons gauge field --- of the form $\mathcal{A}^{(1)}_{\mu} = A_{\mu} (t,\mathbf{x})$.
    \item In $D=3+1$ it becomes a $2$-gauge-form, i.e. an antisymmetric rank-2 tensor --- a.k.a. a Kalb-Ramond gauge field --- of the form $\mathcal{A}^{(2)}_{\mu \nu} = B_{\mu\nu} (t,\mathbf{x})$.
    \item In $D=d+1$ for $d\ge 4$, it is an antisymmetric rank-$p$ tensor $\mathcal{A}^{(p)}_{\mu_{1} \dots \mu_{p}} = C_{\mu_{1}\dots \mu_{p}}(t,\mathbf{x})$.
\end{itemize}
This is a remarkably general result that includes the 1+1D case, for which we initially had no expectations. Interestingly, the topological current expression reduces to the familiar Abelian bosonisation current formula, for which no gauge redundancy is present. Conversely, we can think of the general $D$ expression as a higher-dimensional Abelian bosonisation, also known as \textit{functional bosonisation} \cite{le1996current,fradkin1994fermion,frohlich1995bosonize,burgess1994bosonization,burgess1994bosonization2,schaposnik1995comment,fosco2018functional,schaposnik1998bosonization,banerjee1997501,chanmain13functional,cirio2014tight}. Now, we understand it also describes generalised flux attachment.\\

The statistical ($a_{\mu}$) and topological ($\mathcal{A}_{\mu \nu \lambda \dots}$) gauge fields are rather different objects. Conventional charged matter couples to electromagnetic fields through minimal coupling to a vector potential $\mathbf{p} \rightarrow \mathbf{p} - q \bm{a}$ or, in covariant form, as a source term $\mathcal{J}^{\mu} a_{\mu}\,$. This motivates the assertion that the statistical gauge field is a \textit{physical} gauge field that will couple to matter. It is the relevant degree of freedom that Nature and experiments will care about. As opposed to this, the topological gauge field will be \textit{auxiliary} or \textit{internal}. This implies it will transfer the structural information of flux attachment to the statistical gauge field and disappear in the process. In a way, it is \textit{integrated out} or used to enforce a local constraint. We aim for this to be enough to obtain an explicit flux attachment expression in different dimensions for the statistical gauge field, which allows us to compute its functional form. Once again, we take inspiration from Chern-Simons theory and aim to find a parent topological gauge action that gives us the topological current in Eq.\eqref{eq:generalised_flux} as an equation of motion when the statistical gauge field is minimally coupled to matter. 
Based on the above considerations, we postulate the action 
\begin{equation}
   S  = \int d^{D}x\;\big\{ \,\mathcal{L}_{\text{top.}}[a,\mathcal{A}] - \mathcal{J}^{\mu}a_{\mu}\,\big\}\;\;;\;\;\;\;\; \text{where}\;\;\;\;\;\;\;\;\mathcal{L}_{\text{top.}}[a,\mathcal{A}] = \frac{\kappa}{2\pi}\,\epsilon^{\,\mu \nu \lambda \dots \alpha \beta } \mathcal{A}_{\mu \nu \lambda \dots} \partial_{\alpha} a_{\beta} + \mathcal{L}\,[\mathcal{A}] \;,
\end{equation}
and where we have used $\gamma = \kappa/ (2\pi)\,$. The first term is a Background Field (BF) term that ensures flux attachment when extremising the action for the statistical gauge field. The term $\mathcal{L}\,[\mathcal{A}]$, on the other hand, gives topological dynamics to the topological gauge field. We, however, have not found its expression in all generality as it is changing across dimensions. Nevertheless, we restrict to writing its form in spatial dimensions $d=1$, $d=2$, and $d=3$. Namely
\begin{flalign}\label{eq:chiral_ax}
&1+1D\;:\;\;\;\;\;\;\;\;\;\;\;\;\;\;\;\;\;\;\;\;\;\;\;\;\;\;\;\;\;\;\;\;\;\;\;\;  \mathcal{L}_{\text{top.}}[a_{\mu},\Phi] = \frac{\kappa}{2\pi}\,\epsilon^{\,\mu \nu } \Phi \partial_{\mu} a_{\nu} +  \frac{\kappa}{4\pi}\,\epsilon^{\,01}\partial_{0}\Phi\,\partial_{1}\Phi\;, &\\ \label{eq:mixed_cs}
&2+1D\;:\;\;\;\;\;\;\;\;\;\;\;\;\;\;\;\;\;\;\;\;\;\;\;\;\;\;\;\;\;\;\;\;\;\;\;\; \mathcal{L}_{\text{top.}}[a_{\mu},A_{\mu}] = \frac{\kappa}{2\pi}\,\epsilon^{\,\mu \nu \lambda } A_{\mu} \partial_{\nu} a_{\lambda} +  \frac{\kappa}{4\pi}\, \epsilon^{\,\mu \nu \lambda } A_{\mu} \partial_{\nu} A_{\lambda}\;, &\\ \label{eq:bb_lagran}
&3+1D\;:\;\;\;\;\;\;\;\;\;\;\;\;\;\;\;\;\;\;\;\;  \;\;\;\;\;\;\;\;\;\;\;\;\;\;\;\; \mathcal{L}_{\text{top.}}[a_{\mu},B_{\mu \nu}] = \frac{\kappa}{2\pi}\,\epsilon^{\,\mu \nu \lambda \alpha } B_{\mu \nu} \partial_{\lambda} a_{\alpha} +  \frac{\kappa}{4\pi}\,\epsilon^{\,\mu \nu \lambda \alpha } B_{\mu \nu} B_{\lambda \alpha} \;.
\end{flalign}
whose Euler-Lagrange equations of motion can be computed for both gauge fields, leading to the double identities
\begin{flalign}\label{eq:topo_cons_1d}
&1+1D\;:\;\;\;\;\;\;\;\;\;\;\;\;\;\;\;\;\;\;\;\;\;\;\;\;\;\;\;\;\;\;\;\;\;\;\;\;\mathcal{J}^{\,\mu}_{\,\text{1+1D}} = \frac{\kappa}{2 \pi}\epsilon^{\,\mu \nu} \partial_{\nu} \Phi = \frac{\kappa}{2 \pi}\epsilon^{\,\mu \nu} a_{\nu}\; ,&\\ \label{eq:topo_cons_2d}
&2+1D\;:\;\;\;\;\;\;\;\;\;\;\;\;\;\;\;\;\;\;\;\;\;\;\;\;\;\;\;\;\;\;\;\;\;\;\;\;\mathcal{J}^{\,\mu}_{\,\text{2+1D}} = \frac{\kappa}{2 \pi}\epsilon^{\,\mu \nu \lambda} \partial_{\nu} A_{\lambda} =  \frac{\kappa}{2 \pi} \epsilon^{\,\mu \nu \lambda} \partial_{\nu} a_{\lambda} \;, &\\ \label{eq:topo_cons_3d}
&3+1D\;:\;\;\;\;\;\;\;\;\;\;\;\;\;\;\;\;\;\;\;\;\;\;\;\;\;\;\;\;\;\;\;\;\;\;\;\;\mathcal{J}^{\,\mu}_{\,\text{3+1D}} =  \frac{\kappa}{2 \pi}\epsilon^{\,\mu \nu \lambda \alpha} \partial_{\nu} B_{\lambda \alpha} =  \frac{\kappa}{2 \pi}\epsilon^{\,\mu \nu \lambda \alpha} \partial_{\nu} \partial_{\lambda} a_{\alpha}\,,
\end{flalign}
where the conserved Noether matter current $\partial_{\mu}\,\mathcal{J}^{\mu}(x) = 0$ becomes a topological current defined in terms of the statistical gauge field. These are the covariant expressions for generalised flux attachment or, alternatively, the functional forms for a $\text{U}(1)$ statistical gauge field in $d\le 3$. We can write down these results in vector notation. 
\begin{asidebox}
\noindent
We find the corresponding flux attachment Gauss's laws
\begin{flalign}\label{eq:kink1d}
&1+1D\;:\;\;\;\;\;\;\;\;\;\;\;\;\;\;\;\;\;\;\;\;\;\;\;\;\;\;\;\;\;\;\;\;\;\;\;\;   a_{x} \,(t,x) = \frac{2 \pi}{\kappa}\, n\,(t,x) \;, &\\\label{eq:vortex2d}
&2+1D\;:\;\;\;\;\;\;\;\;\;\;\;\;\;\;\;\;\;\;\;\;\;\;\;\;\;\;\;\;\;\;\;\;\;\;\;\;  b \,(t,\mathbf{x}) = \frac{2 \pi}{\kappa}\, n\,(t,\mathbf{x})\;, &\\ \label{eq:monopole}
&3+1D\;:\;\;\;\;\;\;\;\;\;\;\;\;\;\;\;\;\;\;\;\;\;\;\;\;\;\;\;\;\;\;\;\;\;\;\;\;  \bm{\nabla} \cdot \bm{b}  \,(t,\mathbf{x}) = \frac{2 \pi}{\kappa}\,n \,(t,\mathbf{x})\;,
\end{flalign}
and quantum Hall-like responses
\begin{flalign}
&1+1D\;:\;\;\;\;\;\;\;\;\;\;\;\;\;\;\;\;\;\;\;\;\;\;\;\;\;\;\;\;\;\;\;\;\;\;\;\;    J_{x}\,(t,x) = - \frac{\kappa}{2\pi}\, a_{t}\,(t,x) \;, &\\
&2+1D\;:\;\;\;\;\;\;\;\;\;\;\;\;\;\;\;\;\;\;\;\;\;\;\;\;\;\;\;\;\;\;\;\;\;\;\;\;  \mathbf{J}\,(t,\mathbf{x}) = - \frac{\kappa}{2\pi}\, \hat{\mathbf{e}}_{z}\times \bm{\varepsilon} \,(t,\mathbf{x}) \;, &\\ \label{eq:hall_3d}
&3+1D\;:\;\;\;\;\;\;\;\;\;\;\;\;\;\;\;\;\;\;\;\;\;\;\;\;\;\;\;\;\;\;\;\;\;\;\;\;  \mathbf{J} \,(t,\mathbf{x}) = - \frac{\kappa}{2\pi}\,\big[\partial_{t}\bm{b}\,(t,\mathbf{x}) + \bm{\nabla}\times \bm{\varepsilon}\,(t,\mathbf{x})\big] \;,
\end{flalign}
where $\bm{b} = b^{i} = \epsilon^{\,ijk} \partial_{j} a_{k}$ and $\bm{\varepsilon} = \varepsilon^{i} = f^{0i} = \partial^{0}a^{i} - \partial^{i} a^{0}$ correspond to the ``magnetic'' and ``electric'' fields, respectively. For clarity of notation, we have also used $\mathcal{J}^{\mu} = (n,\mathbf{J})$, $\partial_{i} = \bm{\nabla}$, $\partial_{t} = \partial_{0}$, and $a_{0} = a_{t}\,$.
\end{asidebox}

It is worth noting that we can integrate out the topological gauge field and obtain the effective action in terms of the physical statistical gauge field only $S_{\text{eff}}  = \int d^{D}x\;\big\{ \,\mathcal{L}_{\text{eff}}[a] - \mathcal{J}^{\mu}a_{\mu}\,\big\}$ with effective topological Lagrangian density
\begin{flalign} \label{eq:int_out1d}
&1+1D\;:\;\;\;\;\;\;\;\;\;\;\;\;  \mathcal{L}_{\text{eff}}[a_{\mu}] = \frac{\kappa}{2\pi}\,\epsilon^{\,\mu \nu } \bar{a}\,\partial_{\mu} a_{\nu} +  \frac{\kappa}{4\pi}\,\epsilon^{\,01}a_{0}\, a_{1} = \frac{\kappa}{2\pi}\, \bar{a}\tilde{f} +  \frac{\kappa}{4\pi}\,\epsilon^{\,01}\partial_{0} \bar{a}\, \partial_{1}\bar{a} \;,  &\\
&2+1D\;:\;\;\;\;\;\;\;\;\;\;\;\;  \mathcal{L}_{\text{eff}}[a_{\mu}] =  \frac{\kappa}{4\pi}\, \epsilon^{\,\mu \nu \lambda } a_{\mu} \partial_{\nu} a_{\lambda} = \frac{\kappa}{8\pi}\, \epsilon^{\,\mu \nu \lambda } a_{\mu} f_{\nu\lambda}\;, &\\
&3+1D\;:\;\;\;\;\;\;\;\;\;\;\;\;  \mathcal{L}_{\text{eff}}[a_{\mu}] = \frac{\kappa}{16\pi}\,\epsilon^{\,\mu \nu \lambda \alpha } \partial_{\mu} a_{\nu} \partial_{\lambda} a_{\alpha} = \frac{\kappa}{32\pi}\,f_{\mu\nu} \tilde{f}^{\mu \nu} \;,
\end{flalign}
where we have defined the dual field strengths in 3+1D as $\tilde{f}^{\mu\nu} = \frac{1}{2} \,\epsilon^{\,\mu \nu \lambda \alpha}f_{\lambda \alpha}$ and in 1+1D as $\tilde{f} = \epsilon^{\,\mu \nu}\partial_{\mu} a_{\nu}$. Furthermore, we introduce the scalar field
\begin{equation}
    \bar{a}\,(x) = \epsilon_{\alpha\beta}\int d^{2}x'\, \partial^{\,\alpha}_{x} G\,(x-x')\, a^{\,\beta}(x')\;.
\end{equation}
Hence, we naturally recover the familiar Chern-Simons term for 2+1D and the $\theta$-term for 3+1D. We are inclined to think that this pattern will repeat itself giving rise to higher-dimensional versions of these terms for odd ($2n+1D$) and even ($2nD$) spacetime dimensions respectively. This constitutes a dimensional cascade of topological terms, provided that the  $\theta$-term evaluates to a total derivative of a Chern-Simons term at the boundary. This is also suggestive of a cascade of Callan-Harvey anomaly inflow mechanisms.\\

From the above, we might infer the rough form of a statistical gauge field in general dimensions. It should look like
\begin{equation}\label{eq:current_dep}
    a^{\,\mu} (t,\mathbf{x}) \sim \frac{1}{\kappa}\, \epsilon^{\,\mu \nu \lambda \dots \alpha} \int d^{d}\mathbf{x}'\; \bm{\nabla}_{\mathbf{x}} \,\omega_{\nu \lambda \dots} (\mathbf{x} -\mathbf{x}')\,\mathcal{J}_{\alpha}(t,\mathbf{x}')\;,
\end{equation}
where the $\omega_{\mu \nu \dots} (\mathbf{x})$ is a topological ---  and potentially multivalued --- kernel containing non-smooth points and changing with dimension. This might be compactly expressed as a non-trivial pure gauge of the form $a_{\mu} (t,\mathbf{x}) = \frac{1}{\kappa} \partial_{\mu} \,\Omega \, (t,\mathbf{x})$. It is non-trivial provided $\Omega\, (t,\mathbf{x})$ is a topological defect. We can compare this expression with the constraint \eqref{eq:pure_gauge} we obtained in first-quantised non-covariant formulation. Furthermore, Eq. \eqref{eq:current_dep} tells us that the statistical vector potential in general dimensions is a density-dependent --- or covariant-current-dependent --- gauge field. Hence, it becomes trivial $a_{\mu} = \text{const.}$ in absence of matter.  Notwithstanding, the dependence on matter is, with counted exceptions, non-local. This implies that, even though we might evaluate the field in a region of null covariant current $\mathcal{J}_{\mu}(t,\mathbf{x}) \approx 0$, the statistical gauge field at that same point is $a_{\mu}(t,\mathbf{x}) \not\approx 0$ as it depends on the integrated value of the covariant current throughout the whole spatial manifold we are considering, i.e. all other positions $\mathbf{x}'$ in the system. This is not unusual as it is analogous to retarded potentials or Jefimenko’s equations in conventional electrodynamics \cite{griffiths2023introduction}.


\subsection{Discussion} \label{sec:discsec}
We have come a long and abstract way to end up with relatively compact expressions. We have generalised both flux attachment and statistical transmutation to arbitrary dimensions. The overall picture is that conventional quantum matter can become statistically anomalous if coupled to an Abelian gauge field. The particular form of this gauge field changes with dimension, but the physical mechanism is universal. Let us briefly comment in more detail on the results we just found. \\



In 2+1D we recover the familiar results of conventional flux attachment in the long-wavelength limit of FQHE \cite{zhang1989effective,zhang1995chern}. We now see them in new light as we explicitly introduced a topological gauge field, thus making the parent action a sum of a mixed and a pure Chern-Simons terms. In the FQHE context, the pure Chern-Simons term is found for the statistical gauge field $a_{\mu}$ --- as opposed to Eq. \eqref{eq:mixed_cs} --- , which is considered as an internal degree of freedom to be eliminated in favour of the true electromagnetic gauge potential $A_{\mu}$, which comes in as an initially background field. Upon elimination, the Chern-Simons level or, equivalently, the filling fraction, is fractional. We instead consider the statistical gauge field as a physical field and consider $A_{\mu}$ as a topological gauge field. Flux attachment happens for the topological gauge field and it is then effectively transferred to the statistical gauge field. The Chern-Simons level becomes fractional as in the conventional formulation. This slight modification of the conventional effective field theory for FQHE in 2+1D is our control case, while the other results in other dimensions are new.\\

The 1+1D case is, arguably, the most exotic of all. No Chern-Simons term can be defined in a strict sense. In fact, neither can the magnetic field, as there is no sense of the curl of a vector potential in space, so there is no notion of a magnetic flux unless one considers an annular geometry, i.e. periodic boundary conditions. Yet, Eq. \eqref{eq:kink1d} appears as the remnant of flux attachment in 1+1D. This is, despite the non-existence of a magnetic field, a statistical gauge potential can still be defined. In this case the vector potential becomes not only local, but also linear in density. All the apparent simplicity in the form of the gauge field turns into complexity of a parent action containing the analogue of a dynamical $\theta$ or axion term endowed with chiral dynamics. This appears as a consequence of our construction. One can see that in 1+1D a similar argument to the 3+1D case does not work, as upon integrating out the topological gauge field we do not recover a 1+1D version of a $\theta$-term, but Eq. \eqref{eq:int_out1d}. In fact, a mere $\theta$-vacuum in 1+1D is just a background electric field, there is nothing dynamical about it, so it cannot possibly give rise to flux attachment. This is understood by realising that the axion and BF terms coincide in one spatial dimension so, if not subsidised with further dynamics, the axion field decouples from matter and does not play other role than a constant background.\\

\begin{figure}[h]
\centering
    \includegraphics[width=0.4\textwidth]{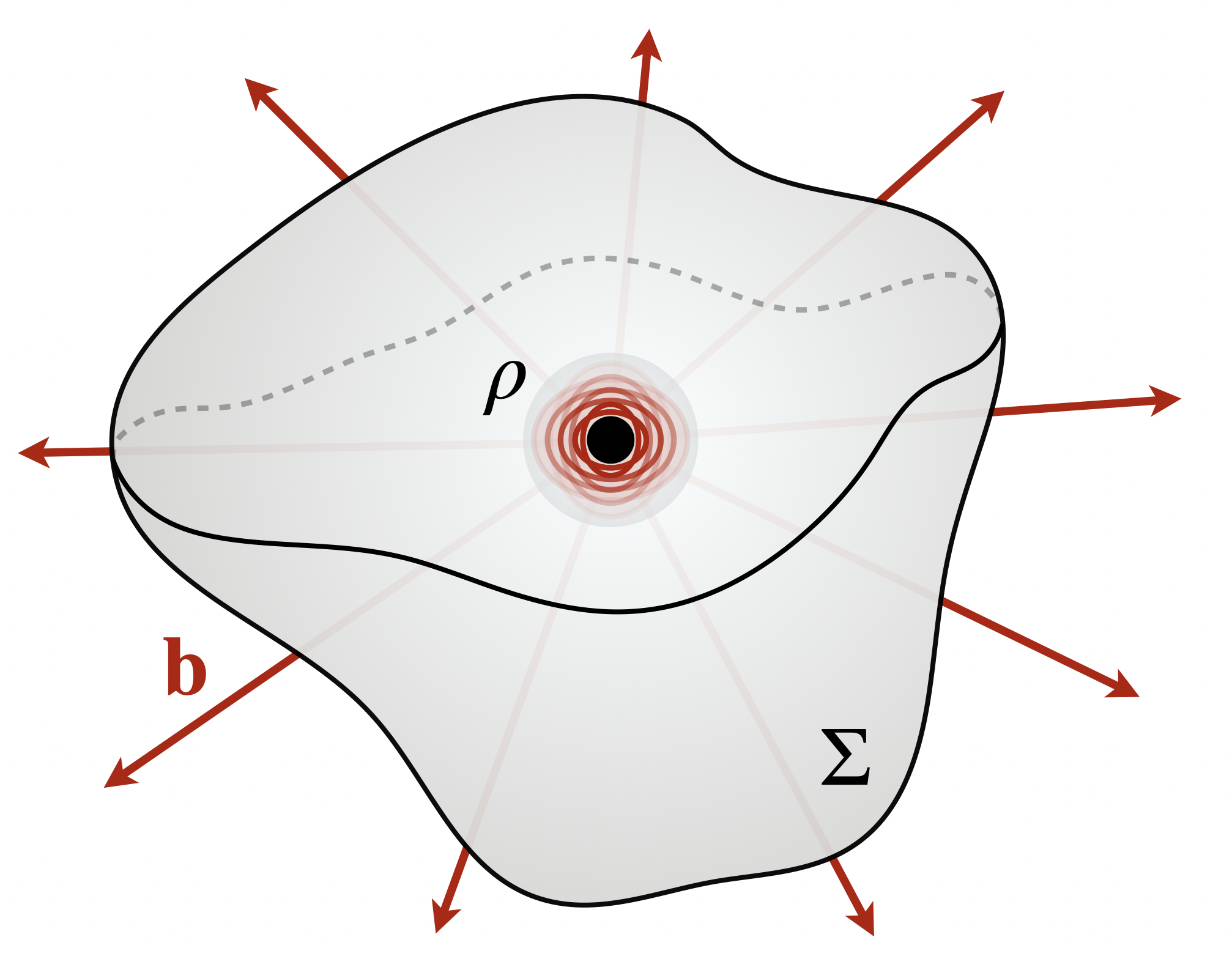}
    \caption{Emergent dyon from flux attachment in three spatial dimensions. A flux-attached electric charge  $\rho (\mathbf{x})$ is placed inside of a closed 3d manifold $\Sigma$. Magnetic flux lines (in dark red) radiate out of the electric charge due to the statistical gauge field profile given by Eq. \eqref{eq:monopole}. The magnetic flux $\Phi^{\Sigma}_{b}$ traversing the boundary $\partial \Sigma$ is proportional to the electric charge $Q_{\Sigma} = \int_{\Sigma} dV\,\rho(\mathbf{x})\,$.}
 \label{fig:ddyon}
\end{figure}

The 3+1D case immediately catches the eye as Eq. \eqref{eq:monopole} is recognised as a variant of the magnetic or Dirac's monopole Gauss's law. In extending electrodynamics to include the existence of hypothetical magnetic monopoles as sources of the magnetic field, we would write $\bm{\nabla}\cdot \mathbf{B}\, (t,\mathbf{x}) \propto \tilde{\rho}_{m}(t,\mathbf{x})$. We say then that the magnetic charge density $\tilde{\rho}_{m}$ acts as a source of magnetic field. In Eq. \eqref{eq:monopole}, however, $n\,(t,\mathbf{x})$ is the electric --- as opposed to magnetic --- charge density. In 2+1D that same electric charge density is proportional to the magnetic field, but in 3+1D is a true source of magnetic field. This means that matter satisfying flux attachment in 3+1D is locally both electric and magnetically charged. Thus, we say that a gauge-matter composite is formed. These electric-magnetic composites have historically been called \textit{dyons} \cite{schwinger1966charge,schwinger1968sources,schwinger1969magnetic,goldhaber1976chargemono} and have relied on the existence of magnetic monopoles. In fact, Witten \cite{witten1979dyons} found that a magnetic monopole in the presence of a $\theta$-term can bind quantised electric charge. This is known as the \textit{Witten effect}. Hence, a magnetic monopole found in a medium with $\theta \neq 0$ is, effectively, a dyon. Here we see the conjugate picture of that which, in the present case, does not rely on the existence of magnetic monopoles. We observe from Eq. \eqref{eq:monopole} that 3+1D flux attachment is a \textit{reciprocal Witten effect} in which a charged particle is endowed with magnetic charge when $\theta \neq 0$. To distinguish the newly found composites from the historically fundamental ones, we might call these \textit{emergent} or \textit{dynamically-generated dyons}. Emergent dyons are, thus, the 3d analogue of Laughlin quasiparticles (or quasiholes). Their effective magnetic charge will be quantised through conventional Dirac quantisation arguments. Integrating Eq. \eqref{eq:monopole} over a 3d spatial volume $\Sigma$ with closed boundary $\partial \Sigma$, we find
\begin{equation}
    Q_{\Sigma}\,(t) = \int_{\Sigma} d^{3}\mathbf{x}\; n\,(t,\mathbf{x}) = \frac{\kappa}{2\pi} \int_{\Sigma} d^{3}\mathbf{x}\;\bm{\nabla}\cdot \bm{b}\,(t,\mathbf{x}) = \frac{\kappa}{2\pi} \int_{\partial \Sigma} d\mathbf{S}\cdot \bm{b}\,(t,\mathbf{x}) \equiv \frac{\kappa}{2\pi}\, \Phi_{b}^{\Sigma}(t)\;,
\end{equation}
which explicitly corresponds to the global expression for flux attachment (see an illustration in Figure \ref{fig:ddyon}). In short, this mechanism generates \textit{dyons without magnetic monopoles}, but by means of charges and statistical gauge fields and through flux attachment in three spatial dimensions. Further evidence of the dyonic behaviour comes from Eq. \eqref{eq:hall_3d} which plays the role of a 3d Hall current, but it also exactly corresponds to Faraday's law of induction in the presence of a magnetic current! This electro-magnetic response is also related in spirit to the usual Topological Magnetoelectric Effect (TME) discussed in axionic topological insulators but explicitly differs in the expressions \cite{wang2015quantised}.

\begin{example}
\noindent
\textbf{Emergent Electro-Magnetic Hadrons}\\

\noindent
Dyons were conceived by Julian Schwinger as an alternative to quarks \cite{schwinger1966charge}. He proposed that Dirac charge quantisation condition imposed by the existence of magnetic monopoles could give rise to bound-states of dyons, namely hadrons, which were magnetically neutral, but had the usual $\frac{1}{3} e$ quantised electric charges. Our framework allows us to conceive a new and emergent view on ``hadrons'' in terms of dyons understood, not as Schwinger's fundamental subatomic entities, but as quasiparticles in, for instance, a 3d quantum fluid. This means a bound-state of two or more emergent dyonic composites, i.e. a composite of composites. These would hypothetically be, in addition to composite bosons/fermions, also electrically-charged (but magnetically neutral)  --- essentially ``electric hadrons'' ---, magnetically-charged (but electrically neutral) --- essentially ``magnetic hadrons'' ---, or a mixture of both, meaning ``electro-magnetic hadrons''.
\end{example}

Some further general remarks are in order. While the present formulation of generalised statistical transmutation is valid for non-relativistic systems, the generalisation of flux attachment in covariant form [Eq. \eqref{eq:generalised_flux}] does not assume any particular matter field. This means that such a formulation is valid for general fermionic/bosonic bare fields regardless of whether they are relativistic or not. In fact, it holds for any covariantly conserved Abelian current. It is likely that the  arguments for generalised statistical transmutation can be extended also to fully relativistic matter field theories.\\

With regards to the proposed parent gauge theories, we find Eqs. \eqref{eq:chiral_ax} to \eqref{eq:bb_lagran} to be in close relation, if not coincident, with Dijkgraaf-Witten-type topological field theories. This unexpected connection deserves further study and will be a subject for future work. Quantisation of the gauge theories is neither discussed nor pursued. The reduction of the gauge action to a mere constraint for the gauge field fully determined by matter is used as a proxy to demand that eventual quantisation of the gauge field must be consistent with that of matter.\\

Another aspect of this extended framework is the possibility of fractional charges. We find that the global flux-charge relation guarantees that in the presence of a quantised magnetic flux, electric charge within the same region will be fractionally quantised by Dirac's argument in $d \ge 2$. We recall that flux is quantised if the region encloses topological defects, and whose presence is ensured by our construction. In $d = 1$ there is no sense of magnetic flux, so the above argument does not hold unless an exception is found if we externally change the topology of the system by considering periodic boundaries in space. Then, the previous arguments are restored since the notion of magnetic flux can be defined on a circle.\\ 

\begin{figure}[h]
\centering
    \includegraphics[width=0.9\textwidth]{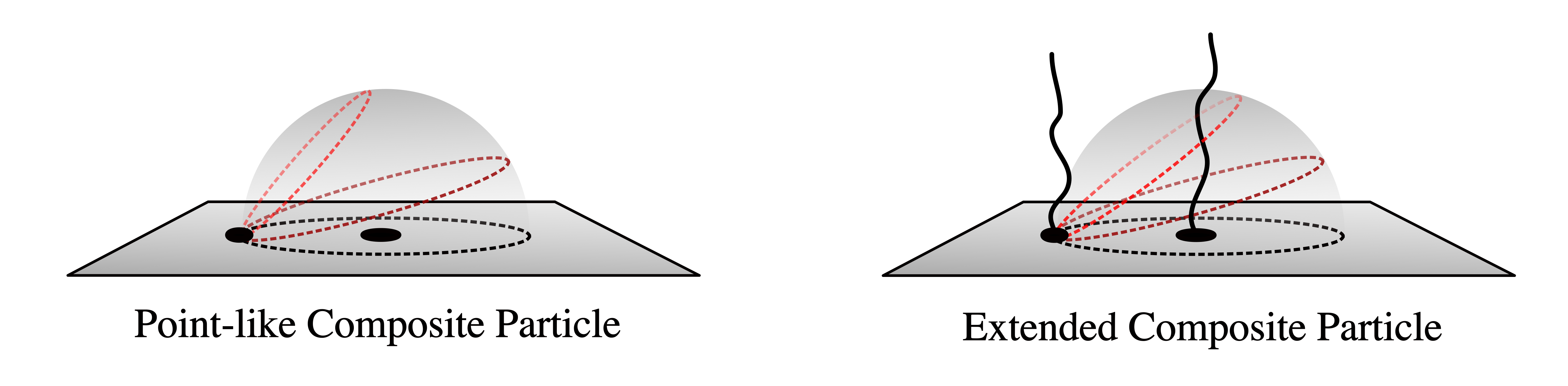}
    \caption{(Left) Retraction argument for a point particle in three-dimensional parameter space. (Right) Obstruction to the retraction map in an analogous scenario when the particles are extended in space. The string may be a line of singularities akin of a Dirac string. }
 \label{fig:string_retract}
\end{figure}

We have shown that statistical transmutation is possible in any dimension, however, we have deliberately prevented ourselves from specifying whether they implied the existence of anyons. In $d\le 2\,$, anyons are allowed according to the common physical lore (see e.g. \cite{leinaas1977theory}). Whereas in $d  = 3\,$, the spin-statistics relation prevents point-like objects from acquiring any statistics that are neither Bose nor Fermi. This is not necessarily true any more if some of the postulates leading to such stringent result are relaxed. In particular, if particles are now extended --- for instance string-like --- objects. The rough topological argument is that for a point-like singular object at $\mathbf{x}_{0}$ in a 2d plane one cannot retract a closed line contour enclosing the defect to a point. In 3d, one can make use of the third spatial dimension to avoid the obstruction, and thus retract to a point. If the defect is now not point-like but string-like (see Figure \ref{fig:string_retract}), retraction is not possible anymore in 3d, but only in 4d and higher. From this heuristic argument, we see the anyons considered in this view will necessarily be extended in higher dimensions. Bose and Fermi statistics and their transmutation are, however, allowed in arbitrary $d$.

\begin{asidebox}
\begin{thm}
 Polyakov's Bose-Fermi correspondence \cite{polyakov1988fermi} in 2+1D, understood as a statistical transmutation performed by a Chern-Simons gauge field, constitutes the incarnation of a physical process that can happen in any dimension.   
\end{thm}

\noindent
\textbf{Caveat}.
Statistical transmutation does \textit{not} imply anyonic statistics. For instance, in 3+1D, homotopical arguments \cite{freedman2011projec} show that in the absence of extended defects, statistics can only be either Bose-like or Fermi-like, meaning there are no anyons. However, statistical transmutation is possible between bosons and fermions. More generally, statistical transmutation can be defined more loosely, to then be constrained by the dimensionality of the system and that of the defects. That is, point-like defects can be lifted in three space dimensions, but string-like defects might not. The possibility of fractional statistics \cite{jian2014layer} will be intrinsically dependent on this.
\end{asidebox}

\subsection{Jordan-Wigner Gauge Dressing and Duality}
We have witnessed the construction of statistical gauge fields that can be written as topologically non-trivial pure gauges. These gauge fields both couple and bind to matter. Upon performing large gauge transformations, statistical gauge fields can be removed at the expense of changing the statistical nature of matter fields. The transformation is a duality that physically connects the bare fields to the homotopically distinct, but equivalent, composite fields. This operation reads as
\begin{equation}
	\Psi_{\,C}\,(\mathbf{x}_{1},\dots,\mathbf{x}_{N}) = e^{\,- i\alpha \,\sum_{m<l}\phi\,(\mathbf{x}_{m},\mathbf{x}_{l})}\, \Psi_{\,B}\,(\mathbf{x}_{1},\dots,\mathbf{x}_{N}) \;,
\end{equation}
where $\phi$ is a disorder scalar field with non-trivial exchange of pairs of labels, i.e. it is topological. In a sense, such an expression is nothing but the $d$-dimensional generalisation of the classic Jordan-Wigner transformation, expressed in continuum and first-quantised language. In second-quantised formalism, we can obtain an equivalent expression for it, namely
\begin{equation}\label{eq:order-disorder}
\hat{\Psi}_{\,C}\,(t,\mathbf{x}) =	e^{-\frac{i}{\hbar} \hat{\Phi}(t,\mathbf{x})} \,\hat{\Psi}_{\,B}\,(t,\mathbf{x}) \;,
\end{equation} 
where $\hat{\Psi}_{\,B}$ and $\hat{\Psi}_{\,C}$ are, respectively, the bare and composite matter field operators, and
\begin{equation}\label{eq:jw-brane}
    \hat{\Phi} \,(t,\mathbf{x}) = \alpha \int d^{d}\mathbf{x}'\;\omega\, (\mathbf{x} - \mathbf{x}') \;\hat{n}\, (t,\mathbf{x}')
\end{equation}
is a real scalar field that sums the contributions of the number density at all positions, weighted by a topological kernel. This reduces to a generalised Jordan-Wigner string object for $d=1$ and a Jordan-Wigner ``membrane'' for $d=2$. Borrowing terminology from string theory, we might call the general case for Eq. \eqref{eq:jw-brane} a $d$-dimensional Jordan-Wigner \textit{brane} (or \textit{JW} $d$-\textit{brane} for short). We can now define a disorder operator $\hat{\mathcal{W}}(t,\mathbf{x}) = \exp{[\frac{i}{\hbar} \hat{\Phi}(t,\mathbf{x})]}$ as the object performing the transformation. This operator creates (removes) topological defects of codimension $d$ in an, otherwise ordered, configuration when applied to a state. Once again, these defects are a kink in $d=1$, a vortex in $d=2$, a monopole in $d=3$, or a general topological soliton in arbitrary $d$. This familiar object, that we see appear often in some embodiment in duality contexts, plays now a clear role. It ``gauge dresses'' the bare matter field and creates a topologically non-trivial composite. We see that this also brings new meaning to the usual Jordan-Wigner transformation, as we know now it is a large gauge transformation connecting bare and composite fields. The conventional Jordan-Wigner transformation is nothing but the remnant of flux attachment and statistical transmutation in 1+1D. Conversely, the Jordan-Wigner transformation in any given dimension is a Bose-Fermi correspondence understood as a particular instance of the composite particle duality, and thus, intimately related to statistical gauge fields and topological field theory.\\

Coming back to Eq. \eqref{eq:order-disorder}, we can call $\hat{\Psi}_{B}$ an \textit{order operator} as an extension of the notion of an order parameter, typically defined as the expectation value of the relevant quantum field. Then, Eq. \eqref{eq:order-disorder} reads symbolically as a recipe:
\begin{asidebox}
\begin{thm}
 Going from Bose to Fermi is actually going from Bare to Composite. A Composite field is a dressed Bare field.   
\end{thm}
\centering
    \textbf{Composite} Operator = \textbf{Disorder} Operator  $\times$  \textbf{Order} Operator  .
\end{asidebox}
\noindent This order-disorder structure has been highlighted in literature since the early works of Kadanoff \cite{kadanoffceva1971}, and has continued to appear recursively \cite{hasenfratz1975fermion,hasenfratz1979puzzling,marino2017quantum,marino1993quantum,kleinert1989gauge,fradkin2017disorder,fradkin2013,fradkin2021quantum,karch2016particle,karch2019arfmain}. We support the view of several authors of the order-disorder structure being the seed of Bose-Fermi dualities. We see that this is not by chance, but the manifestation of flux attachment and statistical transmutation in higher degree of generality. Understood as a recipe it is tempting to conclude that this order-disorder structure is extremely robust, likely beyond our limited study here.\\

Beyond the implications already stated, the transformation in Eq. \eqref{eq:order-disorder} has several peculiarities. Upon evaluation of conjugate matter fields in arbitrary dimensions, we find
\begin{equation}
    \hat{\Psi}^{\dagger}_{\,C} (t,\mathbf{x}')\hat{\Psi}_{\,C}(t,\mathbf{x}) = \hat{\Psi}^{\dagger}_{\,B} (t,\mathbf{x}')\,\hat{\mathcal{W}}(t,\mathbf{x}')\,\hat{\mathcal{W}}^{\dagger}(t,\mathbf{x})\,\hat{\Psi}_{\,B} (t,\mathbf{x})\;.
\end{equation}
We observe that for $\mathbf{x} = \mathbf{x}'\,$, disorder operators evaluate to the identity $\hat{\mathcal{W}}^{\dagger}(\mathbf{x}) \hat{\mathcal{W}}(\mathbf{x}) = \hat{\mathbb{I}}\,$, i.e. opposite defects annihilate. The immediate implication is that the density operator $\hat{n} \,(t,\mathbf{x}) = \hat{\Psi}_{\,B}^{\dagger} (t,\mathbf{x})\,\hat{\Psi}_{\,B}(t,\mathbf{x}) = \hat{\Psi}_{\,C}^{\dagger}(t,\mathbf{x}) \,\hat{\Psi}_{\,C}(t,\mathbf{x})\,$ is preserved by the transformation. Intuitively, as it happens in conventional Jordan-Wigner transformations, the strings cancel each other at the same position. Here, we see that the JW $d$-branes must cancel each other by construction when evaluated at the same position. This means that any operator that can be expressed as local powers of the density operator will be invariant under the mapping. This is important as it includes certain local interactions at the level of the field theoretical Lagrangian or Hamiltonian. Thus, our generalisation of flux attachment, which was proved by systems free of interparticle interaction, holds at the exact same level of rigour if interactions are of the form
\begin{equation}
\hat{\mathcal{H}}_{\,\text{int}} = g_{0}(t,\mathbf{x}) + g_{1}(t,\mathbf{x})\,\hat{n}(t,\mathbf{x}) + g_{2}(t,\mathbf{x}) \,\hat{n}(t,\mathbf{x}) \hat{n}(t,\mathbf{x})  + \,\dots \,+ g_{n}(t,\mathbf{x})\, \big[\hat{n}(t,\mathbf{x})\big]^{n}\;.
\end{equation}
The cancellation of JW $d$-branes also implies that the statistical properties of the composite field reduce to those of the bare field at $\mathbf{x} = \mathbf{x}'$. In other words, similar to what happens with conventional anyons or Jordan-Wigner transformed fields, the composite field will have a \textit{hard} or \textit{soft core} depending on whether the bare field is a fermion or a boson.  Hence, the composite fields can be hard-core or soft-core bosons, fermions or anyons. For $\mathbf{x}\neq \mathbf{x}'\,$, the product $\hat{\mathcal{W}}^{\dagger}(\mathbf{x}') \hat{\mathcal{W}}(\mathbf{x})$ is crucially not equal to the identity. The mapping still holds, but the system is no longer described locally as there are strings connecting these defects, which are spatially separated. The fact that these disorder operators have non-trivial commutation is precisely what allows for the transmutation of statistics.

\section{A Recipe for Novel Topological Quantum Matter}\label{sec:sitelink}

Armed with our knowledge of the composite particle duality we can now introduce models of quantum matter that become \textit{topologically ordered} by construction upon coupling to a statistical gauge field. These models, interacting and highly nonlinear in the bare basis, have a particularly simple and local composite dual description, which allows for exact or simplified solutions. This family of models serves as an illustration and proof-of-principle, meaning it is not limited to those explicitly discussed. Some previously known examples of topologically ordered matter may appear as particular instances of our framework. They can, in principle, exist in $d$ dimensions, although limited explicit computation of the required gauge field restricts our discussion to $d \le 3\,$.

\subsection{In Continuum.} We regard as conventional quantum matter those systems that follow the usual Landau symmetry-breaking paradigm. For non-relativistic $D$-dimensional systems, a Lagrangian density of the form 
\begin{equation}
    \mathcal{L} = i\hbar\, \hat{\Psi}^{\dagger}(x) \partial_{t} \hat{\Psi}(x) - \frac{\hbar^{2}}{2m} \bm{\nabla} \hat{\Psi}^{\dagger}(x)\cdot \bm{\nabla}\hat{\Psi}(x) - V(x)\, \hat{\Psi}^{\dagger} (x) \hat{\Psi} (x) - U(x)\,\hat{\Psi}^{\dagger}(x) \hat{\Psi}^{\dagger}(x)\hat{\Psi}(x) \hat{\Psi}(x)
\end{equation}
with notation $x \equiv (t,\mathbf{x})$ and satisfying (anti)commutation relations
\begin{flalign}\label{eq:comm1}
    &\big[ \hat{\Psi}(x), \hat{\Psi}(x')\big]_{\,\mp} = \big[ \hat{\Psi}^{\dagger}(x), \hat{\Psi}^{\dagger}(x')\big]_{\,\mp} = 0 \;\\ \label{eq:comm2} 
    & \big[ \hat{\Psi}(x), \hat{\Psi}^{\dagger}(x')\big]_{\,\mp} = \delta^{(D)} (x - x')\;,
\end{flalign}
describes a conventional Bose or Fermi quantum fluid interacting locally in spacetime. Matter fields in this theory interact pairwise via a short-range potential of the form $U(x,x') = U(x)\,\delta^{(D)}(x - x')$. Three-body or higher-body interactions are neglected for simplicity. Such a model describes not one, but a family of systems, approximate incarnations of which can be found in atomic condensates or electron fluids with screened interactions. This is clearly conventional quantum matter. Our study of the composite particle duality suggests that the mere coupling to a statistical gauge field, drastically alters the system and automatically turns it into a topological quantum fluid, as in topologically ordered and statistically anomalous. This new family of models will be described by the Lagrangian density
\begin{equation}\label{eq:lagrangian_qfluid}
\begin{split}
    \mathcal{L} &= i\hbar\, \hat{\Psi}^{\dagger}(x) \,\Big(\partial_{t} - \frac{i}{\hbar} \,\hat{a}_{0} (x)\Big)\,\hat{\Psi}(x) - \frac{\hbar^{2}}{2m}
    \,\Big[\Big(\bm{\nabla} - \frac{i}{\hbar}\, \hat{\bm{a}}(x)\Big) \,\hat{\Psi}(x)\Big]^{\dagger}\cdot \Big(\bm{\nabla} -\frac{i}{\hbar}\, \hat{\bm{a}}(x)\Big)\,\hat{\Psi}(x) \\ 
    &- V(x)\, \hat{\Psi}^{\dagger} (x) \hat{\Psi} (x) - U(x)\,\hat{\Psi}^{\dagger}(x) \hat{\Psi}^{\dagger}(x)\hat{\Psi}(x) \hat{\Psi}(x)
\end{split} 
\end{equation}
with the statistical gauge field $\hat{a}_{\mu}(x)$ satisfying the corresponding $D$-dimensional flux attachment constraint in Eqs. (\ref{eq:topo_cons_1d}--\ref{eq:topo_cons_3d}). Hence, the statistical gauge field will in general be a non-local function of the matter current $\hat{a}_{\mu} = \hat{a}_{\mu}[\hat{J}^{\mu}]$. This implies that the previous theory can be encoded or re-expressed in matter as an interacting quantum matter theory, meaning Eq. \eqref{eq:lagrangian_qfluid} can be rewritten as 
\begin{flalign}
 \mathcal{L} &= i\hbar\, \hat{\Psi}^{\dagger}(x) \partial_{t} \hat{\Psi}(x) - \frac{\hbar^{2}}{2m} \bm{\nabla} \hat{\Psi}^{\dagger}(x)\cdot \bm{\nabla}\hat{\Psi}(x) - V(x)\, \hat{\Psi}^{\dagger} (x) \hat{\Psi} (x)
 - U(x)\,\hat{\Psi}^{\dagger}(x) \hat{\Psi}^{\dagger}(x)\hat{\Psi}(x) \hat{\Psi}(x) - \hat{J}^{\mu}(x)\, \hat{a}_{\mu}(x)
\end{flalign}
where the last term should be thought of as a \textit{statistical interaction}. In other words, it is the normal minimal coupling, but since the gauge field has no free dynamics it can be fully encoded in matter and appear indistinguishable from conventional interactions \footnote{This should be considered as opposed to more conventional Maxwell-like theories with propagating modes that exist even in the absence of matter, namely photons. There are no free photons in the theories proposed.}. We expect statistical interactions to significantly alter the groundstate of the system as well as its dynamics and excitations. Even at weak couplings nonlinearities will develop and are likely to dominate the dynamics. As a consequence, we expect exotic soliton solutions for the matter field $\hat{\Psi}$ to be found. Quantum fluids with this kind of interactions will be topological and accept a dual description
\begin{flalign}
    \tilde{\mathcal{L}}_{C} &= i\hbar\, \hat{\tilde{\Psi}}_{C}^{\dagger}(x) \partial_{t} \hat{\tilde{\Psi}}_{C}(x) - \frac{\hbar^{2}}{2m} \bm{\nabla} \hat{\tilde{\Psi}}_{C}^{\dagger}(x)\cdot \bm{\nabla}\hat{\tilde{\Psi}}_{C}(x) - V(x)\,
    \hat{\tilde{\Psi}}_{C}^{\dagger} (x) \hat{\tilde{\Psi}}_{C} (x) 
    - U(x)\,\hat{\tilde{\Psi}}_{C}^{\dagger}(x) \hat{\tilde{\Psi}}_{C}^{\dagger}(x)\hat{\tilde{\Psi}}_{C}(x) \hat{\tilde{\Psi}}_{C}(x)
\end{flalign}
in terms of a composite field $\hat{\tilde{\Psi}}_{C}$ that will no longer satisfy relations \eqref{eq:comm1} and \eqref{eq:comm2}. 

\subsection{On the Lattice.} 
\begin{figure}[h]
\centering
    \includegraphics[width=0.7\textwidth]{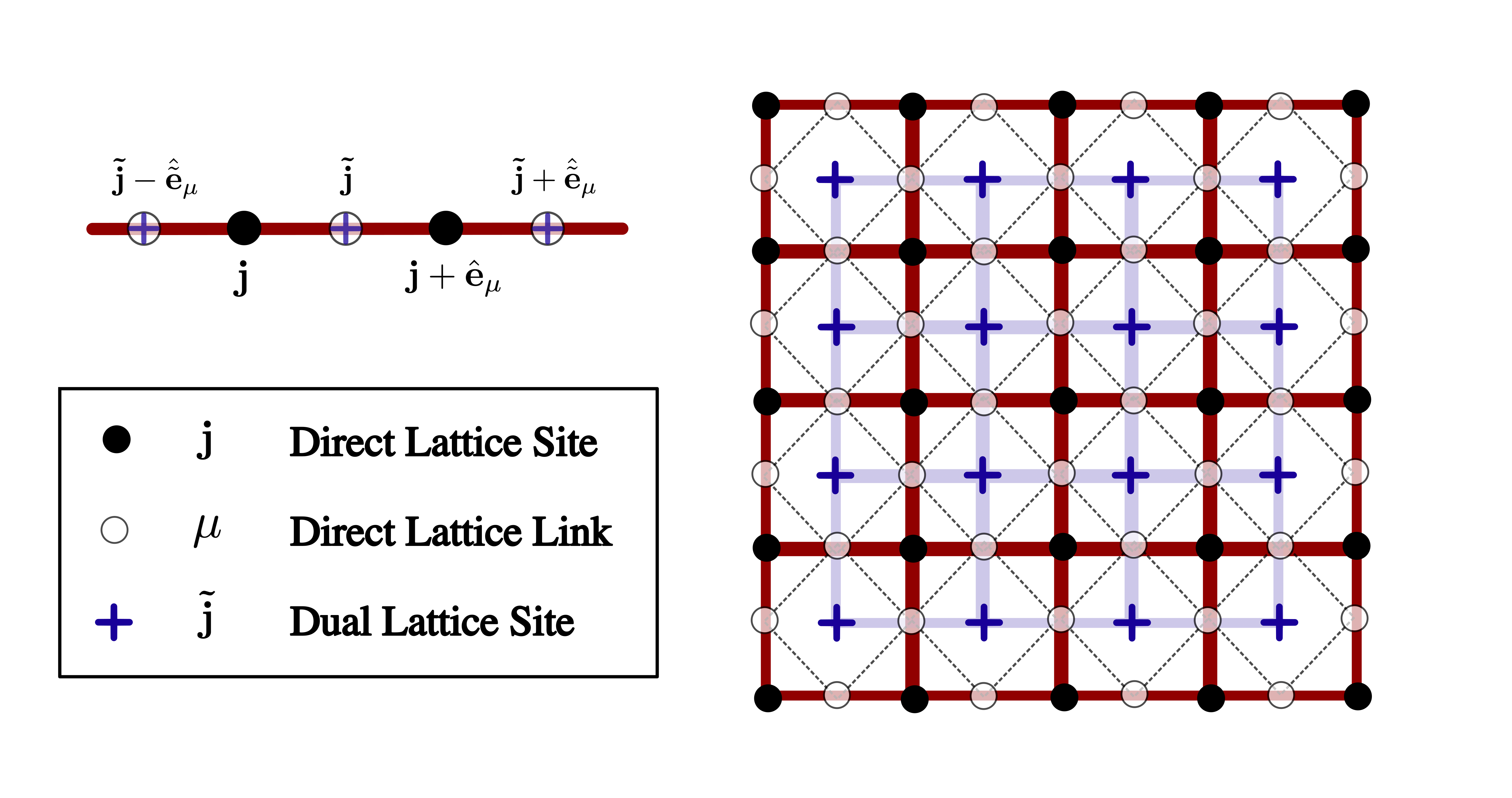}
    \caption{Direct (red) and dual (blue) bipartite lattices in 1d and 2d.}
    \label{fig:lattice_dual}
\end{figure}

The previously introduced model and arguments can also be considered in a lattice setup, suitable for the description of a vast portion of quantum solids. For simplicity, we restrict ourselves to work in the tight-binding approximation for a regular lattice. In this case, we consider the $d$-dimensional lattice Hamiltonian 
\begin{equation}\label{eq:quant_solid}   \hat{H} = \sum_{\mathbf{j}=1}^{N}\Big[\sum_{\mu=1}^{\alpha}\Big(\hat{c}^{\dagger}_{\mathbf{j}} \,J_{\mathbf{j},\mathbf{j} +\hat{\mathbf{e}}_{\mu}}(t)\,\hat{c}_{\mathbf{j}+\hat{\mathbf{e}}_{\mu}} + \text{H.c.} \Big) + V_{\mathbf{j}} (t)\,\hat{n}_{\mathbf{j}} + U_{\mathbf{j}} (t)\,\hat{n}_{\mathbf{j}} \hat{n}_{\mathbf{j}}  + \sum_{\mu=1}^{\alpha} \Big(\hat{n}_{\mathbf{j}}\, W_{\mathbf{j},\mathbf{j} +\hat{\mathbf{e}}_{\mu}} (t)\,\hat{n}_{\mathbf{j} +\hat{\mathbf{e}}_{\mu}} + \text{H.c.}\Big)\Big]
\end{equation}
with (anti)commutation relations
\begin{flalign}
    &\big[ \hat{c}_{\mathbf{i}}, \hat{c}_{\mathbf{j}}\big]_{\,\mp} = \big[ \hat{c}^{\dagger}_{\mathbf{i}}, \hat{c}^{\dagger}_{\mathbf{j}}\big]_{\,\mp} = 0 \;\\ 
    & \big[ \hat{c}_{\mathbf{i}}, \hat{c}^{\dagger}_{\mathbf{j}}\big]_{\,\mp} = \delta_{\,\mathbf{ij}}\;,
\end{flalign}
where $\mathbf{j}$ is the lattice coordinate, $\alpha$ is the number of nearest neighbours and $\hat{\mathbf{e}}_{\mu}$ is a unit vector connecting neighbouring sites along direction $\mu$. This model describes spinless bosons or fermions in a periodic lattice which are allowed to tunnel between neighbouring sites with rate $J$, that experience an on-site potential $V$, and that can interact on-site with strength $U$, or between neighbouring sites with $W$. Here again, we might couple the system to a statistical gauge field. We can do this explicitly by working in the temporal gauge $a_{0}= 0$ and minimally coupling the vector potential. The coupling is achieved performing Peierls substitution, yielding
\begin{equation}\label{eq:extend_hubbard}
    J_{\mathbf{j},\mathbf{j} +\hat{\mathbf{e}}_{\mu}}(t)\longrightarrow J_{\mathbf{j},\mathbf{j} +\hat{\mathbf{e}}_{\mu}}(t)\;\hat{\mathcal{U}}_{\mu}(\mathbf{j})\;,
\end{equation}
where the operator $\hat{\mathcal{U}}_{\mu}$ is a \textit{parallel transporter}, also known as a \textit{Wilson line} or \textit{link variable}, over link $\mu$ of the lattice. For an Abelian group we define it as
\begin{equation}
\hat{\mathcal{U}}_{\mu}(\mathbf{j}) \equiv e^{i\hat{a}_{\mu} (\mathbf{j})}\;\;\;\;\;\;\;\;\;\;\text{and}\;\;\;\;\;\;\;\;\;\;\;\; \hat{a}_{\mu}(\mathbf{j}) \equiv \hat{a}\,(\mathbf{j},\mathbf{j} +\hat{\mathbf{e}}_{\mu}) = \frac{1}{\hbar} \int_{\mathbf{j}}^{\mathbf{j} +\hat{\mathbf{e}}_{\mu}} d\mathbf{x}\cdot\hat{\bm{a}}\,(t,\mathbf{x})\;,
\end{equation}
where $\hat{a}_{\mu}$ is the statistical vector potential and appears as a $\text{U}(1)$ operator-valued Peierls phase. In fact, it is a density-dependent Peierls phase.  It is important to notice that matter fields live on the direct lattice sites $\mathbf{j}$, gauge fields live on the links of the lattice $\mu$, and disorder operators live on the sites of the dual lattice $\mathbf{\tilde{\mathbf{j}}}$. Links $\mu$ and dual sites $\mathbf{\tilde{\mathbf{j}}}$ are coincident for $d=1$, but differ substantially in $d>1$ (see Figure \ref{fig:lattice_dual}).
The form of the gauge potential can be explicitly implemented by the flux attachment constraints in the corresponding dimension. Alternatively, one can implement these constraints directly on the lattice, so that the model becomes a \textit{lattice gauge theory}. This is defining a Gauss's law operator $\hat{G}_{\mathbf{j}}$ restricting the Hilbert space of the system to a physical subspace $\mathscr{H}_{\,\text{Phys}} \subset \mathscr{H}$ obeying the local flux attachment constraint 
\begin{equation}\label{eq:gauss_flux}
    \hat{G}_{\mathbf{j}} \ket{\text{Phys}} \equiv \Big[ \hat{\beta}_{\,\mathbf{j}} - \exp{\big(\gamma \hat{n}_{\mathbf{j}}\big)} \Big] \,\ket{\text{Phys}} = 0\;,
\end{equation}
where $\gamma$ is the statistical factor. Note that $\ket{\text{Phys}}$ represent the physical states of the system and that $[\hat{G}_{\mathbf{j}}\,, \hat{H}] = 0$ is satisfied at any point in spacetime. The operator $\hat{\beta}_{\mathbf{j}}$ is a scalar field that changes with dimension and is a function of the statistical gauge fields. It plays the role of a magnetic flux on the dual lattice (e.g. blue lines in Figure \ref{fig:lattice_dual}) or, alternatively, in the links lattice (e.g. dashed lines in Figure \ref{fig:lattice_dual}). We shall call it the \textit{statistical flux operator} \footnote{It is important to distinguish this operator from conventional disorder (i.e. kink, vortex, monopole) operators we find in conventional literature.} from now on. Thus, it is defined with local reference to the links $\mu$ coming out of the direct lattice site $\mathbf{j}$. More explicitly, it is the product of links associated to a given direct lattice site
\begin{equation}
    \hat{\beta}_{\,\mathbf{j}} = \prod_{\braket{\mu}} \hat{\mathcal{U}}_{\mu}(\mathbf{j})\;,
\end{equation}
where $\braket{\bullet}$ signals that the product is taken for all the link variables associated to location $\mathbf{j}$. It is the lattice equivalent to computing the magnetic flux centred at $\mathbf{j}$, which ensures the charge-flux binding. We refer the reader to Appendix \ref{sec:discsec} for an intuitive discussion of this object. That is with the exception of $d=1$, provided no magnetic flux can be defined and the statistical flux operator reduces to $\hat{\beta}_{\,\mathbf{j}} = \hat{\mathcal{U}}_{\mu}(\mathbf{j})$ in the most convenient gauge. The discrete version of the composite dual mapping \eqref{eq:order-disorder} at a given lattice site becomes a $d$-dimensional Jordan-Wigner transformation
\begin{equation}\label{eq:dual_discrete}
    \hat{f}_{\mathbf{j}}\,(\Gamma) = \hat{\mathcal{W}}^{\dagger}_{\tilde{\mathbf{j}}} (\Gamma)\,\hat{c}_{\mathbf{j}}\;,\;\;\;\;\;\;\;\text{with}\;\;\;\;\;\;\;\;\;\;\;\; \hat{\mathcal{W}}^{\dagger}_{\tilde{\mathbf{j}}} (\Gamma)\,\hat{\mathcal{W}}_{\,\tilde{\mathbf{j}}}\,(\Gamma) = \hat{\mathbb{I}}\;,
\end{equation}
where $\Gamma$ denotes a \textit{'t Hooft magnetic brane}, which is a topological defect of appropriate codimension inherited by the composite field. The number density operator is locally invariant under large gauge transformations like Eq. \eqref{eq:dual_discrete}, yielding $\hat{n}_{\mathbf{j}} = \hat{c}^{\dagger}_{\mathbf{j}} \hat{c}_{\mathbf{j}} = \hat{f}^{\dagger}_{\mathbf{j}} \hat{f}_{\mathbf{j}}\,$ as expected from continuum. Provided most of the terms in the model \eqref{eq:quant_solid} --- minimally coupled through Eq. \eqref{eq:extend_hubbard} and constrained by Eq. \eqref{eq:gauss_flux} --- are expressed in powers of the number density, only the kinetic term will be altered in the composite dual picture. \\

With all the above, the family of interacting tight-binding models in $d$ spatial dimensions \eqref{eq:quant_solid} admits a simple composite dual description
\begin{equation}  \label{eq:compduallattice}
\hat{\tilde{H}}_{C} = \sum_{\mathbf{j}=1}^{N}\Big[\sum_{\mu=1}^{\alpha}\Big(\hat{f}^{\dagger}_{\mathbf{j}}(\Gamma) \,J_{\mathbf{j},\mathbf{j} +\hat{\mathbf{e}}_{\mu}}(t)\,\hat{f}_{\mathbf{j}+\hat{\mathbf{e}}_{\mu}}(\Gamma) + \text{H.c.} \Big) + V_{\mathbf{j}} (t)\,\hat{n}_{\mathbf{j}} + U_{\mathbf{j}} (t)\,\hat{n}_{\mathbf{j}} \hat{n}_{\mathbf{j}}  + \sum_{\mu=1}^{\alpha} \Big(\hat{n}_{\mathbf{j}}\, W_{\mathbf{j},\mathbf{j} +\hat{\mathbf{e}}_{\mu}} (t)\,\hat{n}_{\mathbf{j} +\hat{\mathbf{e}}_{\mu}} + \text{H.c.}\Big)\Big]\,.
\end{equation}
Notice that the composite dual of \textit{any} spinless tight-binding system can always be obtained by performing the transformation \eqref{eq:dual_discrete}, provided it is an operator identity. This statement is nothing but that of a, now generalised, Jordan-Wigner transformation being always valid, although not always useful. We see that, despite the transformation being extremely non-local, the dual model is, in general, reasonably local if no anomalous terms \footnote{An example can be Cooper-pairing like terms such as $\hat{c}_{\mathbf{j}} \hat{c}_{\mathbf{j}}$ or $\hat{c}^{\dagger}_{\mathbf{j}} \hat{c}^{\dagger}_{\mathbf{j}}$.} are present due to cancellation of JW $d$-branes. Note that this is a type of duality in matter, meaning that both the bare and the composite fields \textit{live} on the same point of the lattice, contrary to classic electric-magnetic, Kramers-Wannier or order-disorder dual models, which move between the direct and dual spatial lattices. The latter notion, however, is contained in the flux attachment constraint \eqref{eq:gauss_flux}, provided gauge fluxes are defined on the dual lattice but with reference to matter living on the direct lattice. The dual lattice, and correspondingly the notion of magnetic flux, change with dimension. All composite models are lattice gauge theories in disguise. This includes the usual Jordan-Wigner transformed systems in 1d, as illustrated in Appendix \ref{sec:hidden}.\\

The family of proposed models are, by construction, gauge theories and topologically ordered. However, as previously stated in the continuum case, they can also be fully encoded in matter. This will lead to non-local interactions in high dimensions, but can reduce to standard local or semi-local interactions in low dimensions, making the system indistinguishable from a conventional model of interacting matter fields. Hence, a new realm of topological models is expected, and well-known examples might emerge as particular cases of this proposal. Particularly intuitive limits of the above families [Eq. \eqref{eq:compduallattice}] are the spinless anyon-Hubbard models in 1d ($\alpha=1$) and 2d ($\alpha = 2$) in either their hard-core or soft-core avatars, which correspond to the simple case of $U=W=0$ and $J_{\mathbf{j},\mathbf{j} +\hat{\mathbf{e}}_{\mu}} = J = \text{const.}$ in the composite dual basis. In fact, examples of these particular cases might be hidden in real materials.

\section{Closing Remarks}
\label{sec:summary}

The quantum statistics of matter can be altered in interacting many-body systems by means of statistical gauge fields. We have extended this familiar notion in the context of FQHE in 2+1D to a broader framework in arbitrary dimensions. We have introduced and discussed a composite particle duality, which provides an exact mapping from conventional to statistically transmuted quantum matter. The duality gives as output the type of gauge field and transformation needed to turn an ordinary system into a statistically exotic one. To be more concrete, we provide a recipe and a physical mechanism to dynamically turn ordinary (Landau-like) quantum solids and fluids into topologically-ordered (beyond-Landau) quantum matter. This enables us to write down entire families of interacting models in $d$ spatial dimensions that have a simple composite dual model but can be expressed in terms of conventional systems coupled to a fluctuating statistical gauge field. The anyon-Hubbard models, anyonic Lieb-Liniger fluids or certain FQH fluids are found as simple instances of this prescription. This provides a unifying framework to understand and modify the statistical properties of phases of matter across dimensions.\\ 

Our formulation predicts the existence of emergent dyons out of flux attachment in 3+1D without invoking the notion of magnetic monopoles, i.e. dyons created as a result of a certain gauge field profile being attached to electric charges. In 1+1D we show that statistical transmutation driven by a synthetic gauge field is still allowed, even though the notion of flux attachment is no longer possible. Some phenomenological consequences of the latter statement have been investigated in Ref. \cite{valenti2024dual}. For particular cases under study in a lattice setting, we identify the statistical flux operator as the key element to enforce the generalised flux attachment constraint via Gauss's law on lattice sites.\\

Statistical gauge fields are found to be the key to change the identity of matter. The form of these gauge fields is set by a generalised flux attachment law acting as a local constraint, and whose explicit form changes depending on dimensionality. These gauge fields are peculiar as they admit a re-expression solely in terms of matter, so there are no free propagating ``photons'' in the absence of matter. In fact, since statistical gauge fields can be rewritten as matter interaction terms, they might be indistinguishable from them. Sufficiently strong interactions are believed to be a necessary condition for the emergence of topological order. The results presented here are expected to describe the low-energy effective limit of quantum phases of matter, for which some of these interactions might be relabeled as statistical gauge terms.\\

Ability to engineer and tune these interactions or gauge fields would allow manipulating and transforming ordinary quantum systems into exotic quantum matter. Recent experiments in 1+1D, performed at ICFO \cite{frolian2022realising} for a Bose-Einstein condensate and at Harvard \cite{kwan2023realization} with ultracold bosons in an optical lattice, have already shown that said manipulation is attainable with the current state-of-the-art technology. We expect these experiments to provide new avenues for the understanding of topological phases of matter and the implementation of topological gauge theories. We believe the composite particle duality \cite{valenti2023composite} is a consistent and insightful framework for a phenomenological understanding of those systems.

\subsection{Open questions}

Statistical transmutation is a purely quantum-mechanical effect, as the notion of quantum statistics does not have a classical counterpart. This means that statistical gauge fields lose their interpretation as dressing agents altering quantum numbers when in a classical setup. However, the notion of flux attachment and its generalisation is not necessarily related to quantum mechanics. Hence, we can indeed consider classical flux attachment, in which matter fields are bound to gauge field defects and create unusual composites, but one should adapt the interpretation. Classical systems are essentially determined by their dynamics, their equations of motion. The minimal coupling of matter to gauge fields that are functions of matter densities and currents --- as the ones discussed in this work ---  will create nonlinearities in \textit{all} cases. These nonlinearities arise from self-interaction or self-interference, and will grow in range as we consider higher dimensional systems. At least for the Abelian cases considered, these gauge fields can be completely eliminated in favour of matter. There is no guarantee this should be true if one extends the composite particle duality or generalised flux attachment to non-Abelian gauge fields. It would be interesting to consider examples of the latter in low dimensions and get an intuition for what features can the richer internal gauge structure offer. We expect the differences to be analogous, for instance, to those between the Dirac and the 't Hooft-Polyakov monopoles. In a quantum context, and in the models in which it is allowed, this possibility would give rise to non-Abelian anyons for the composites. This, however, remains conjectural at this stage.\\

We should also take the opportunity to comment on the comparison with literature. We find the exciting results reported by Valiente \cite{valiente2020long,valiente2021short} to be an extension of the ones discussed here in 1+1D and, for that matter, also of conventional flux attachment \cite{wilckez1982flux,zhang1995chern} in 2+1D. The starting point of the analysis is based on the use of Shirokov's algebra, which provides a rigorous mathematical treatment of the ill-defined points $\mathbf{x} = \mathbf{x}'$ in the formalism and avoids the rather artificial definitions of hard and soft cores. In other words, it is a potential way to include (or not) Pauli's exclusion principle in the mathematical formalism. At this point it is not exactly clear what would be the phenomenological relevance of that.\\

Following on the connections with established literature, we believe that Bose-Fermi dualities can, for the most part, be understood as composite particle dualities. It is tempting to conjecture on the relation with dualities of Kramers-Wannier type, which are Bose-Bose. There, topological gauge constraints appear also as a key ingredient \cite{savit1980duality}, and admit order-disorder interpretations or particle-soliton correspondences involving gauge fields.  Similarities between FQHE-like and $\mathbb{Z}_{N}$ models with $\theta-$terms were acknowledged in Ref. \cite{shapere1989self} with regards to the formation of an electric-magnetic composite condensate. Further evidence from this potential relation, seems to come from the connection with the $\mathbb{Z}_{N}$ Dijkgraaf-Witten type theories discussed in Ref. \cite{kapustin2014coupling}.
Further study should clarify this connection.\\

Finally, if we are to make any claim about the duality studied here being fundamental and ubiquitous, we should aim to extend all the previous discussion to relativistic systems. In the last decade, \textit{duality webs} have appeared in 1+1D \cite{karch2019arfmain}, 2+1D \cite{seiberg2016duality,karch2016particlevortex} and 3+1D \cite{murugan20214d} as, in part, an implicit extension of the concept of statistical transmutation. This is not surprising, provided that the initial duality that started this discussion, that of Polyakov \cite{polyakov1988fermi}, was manifestly relativistic. In fact, Polyakov's Bose-Fermi duality was extended to 3+1D \cite{fustero1989transmute} shortly after its initial proposal. Should our formalism generalise, it should capture all the above notions. In practice, the generalisation to a complex Klein-Gordon field is trivial. However, extending this framework to real scalars and Dirac spinors is clearly not.

\begin{acknowledgments}
We warmly thank A. Celi, B. Bakkali-Hassani, J. Kwan, J. Pachos, G. Palumbo and L. Tarruell for useful discussions. G.V-R. acknowledges financial support from EPSRC CM-CDT Grant No. EP/L015110/1 and the Naquidis Center for Quantum Technologies. 
\end{acknowledgments}

\clearpage
\newpage
\appendix

\centerline{\large \bf Supplementary Information}
\vspace{0.25cm}

\centerline{Gerard Valentí-Rojas, Joel Priestley, Patrik Öhberg}

\section{Disorder fields across dimensions}\label{sec:snippets}
In order to see the previous concepts at work, let us compute the statistical pure gauge scalar function. In particular, we discuss the case of linear exchange $\phi_{ab} = \beta_{1} + \beta_{2}\, \phi_{ba}$ of labels. We postulate that the $d$-dimensional angular phase of the $d$-dimensional vector $\mathbf{\tilde{x}}_{ab} = \mathbf{x}_{a} - \mathbf{x}_{b}$ with respect to some arbitrary reference $\hat{\text{e}}_{\mathbf{n}}$ direction, is parametrised by functions with the aforementioned exchange property. We want to compare this picture with the field theoretical intuition from generalised flux attachment.\\

\paragraph{One spatial dimension.}
When particles move on a line, there are only two possible choices for direction, left and right. Should one define the analogue of an angle variable or phase given a reference direction, we would find that the only allowed angles are $0$ or $\pi\,$. There is an intrinsic $\mathbb{Z}_{2}$ symmetry and the relevant homotopy group is $\pi_{0}(\mathbb{Z}_{2})$. This degree of freedom can be encoded in a sign ($\pm$) or a binary variable ($0$ or $1$). Hence, a natural choice we can make is the Heaviside step function $\phi_{ml} = \Theta\, (x_{m} - x_{l}) = 1 - \Theta\, (x_{l} - x_{m}) =1 - \phi_{lm}\,$. The corresponding statistical factor can then be obtained explicitly from Eq. \eqref{eq:exchange}, yielding $\gamma = \alpha \sum_{m<l} \text{sgn}\,(x_{m} - x_{l})\,$.  The statistical vector potential is computed as
\begin{flalign}\label{eq:vector_1d}
	&a_{x} \,(x_{i}) = \alpha\,\nabla_{x_{i}} \Phi\,(x_{1},\dots,x_{i},\dots,x_{N}) =\alpha\, \sum_{a<b=1}^{N} \nabla_{x_{i}} \phi_{ab} = \alpha\, \Big(\sum_{b>i} \nabla_{x_{i}} \phi_{ib} + \sum_{a<i} \nabla_{x_{i}} \phi_{ai} \Big) \\
	&= \alpha\, \Big[\sum_{b>i} \delta\,(x_{i} - x_{b}) - \sum_{a<i} \delta\,(x_{a} - x_{i}) \Big] 
	= \alpha\,\sum_{j \neq i} \text{sgn}\,(j-i)\;\delta\,(x_{i} - x_{j}) \equiv \alpha\,n\,(x_{i})\;,
\end{flalign}
where we have used properties $\delta\,(x_{i}) = \nabla_{x_{i}}\Theta\,(x_{i}) $ and $\delta\,(x_{i} - x_{j}) = \delta\,(x_{j} - x_{i})\,$. Notice that the definition for the density is of the form $n(x_{i}) = \sum_{j\neq i} q_{j}\,\delta (x_{i} - x_{j})$ with a charge $q$ that depends not on label position, but on  label ordering. It is with this example that we realise that the phase $\Phi$ acquires the form of a generalised Jordan-Wigner string in which the topological content and a certain nonlocality are encoded. The gauge transformation becomes
\begin{equation}
	\Psi\,(x_{1},\dots,x_{N}) = e^{\,i\alpha \,\sum_{m<l} \Theta (x_{m} - x_{l})}\, \Psi_{c}\,(x_{1},\dots,x_{N}) \;.
\end{equation}

By analogy with the Chern-Simons case, it is expected that some topological gauge theory will naturally give rise to such form of $\Phi$. Preserving the same structure, we see that the field theoretical constraint analogous to Eq. \eqref{eq:vector_1d} would be 
\begin{equation}
	\hat{a}_{x}\,(t,x) = \alpha \,\nabla_{x}\, \hat{\Phi}\,(t,x) = \alpha\,\hat{n}\,(t,x)
\end{equation}
with
\begin{equation}
	\hat{\Phi}\,(t,x) = \int_{-\infty}^{\infty} \text{d}x'\;\Theta\,(x- x')\,\hat{n}\,(t,x') = \int_{-\infty}^{x} \text{d}x'\;\hat{n}\,(t,x')\;,
\end{equation}
being the kink or Jordan-Wigner string. The transformation then becomes
\begin{equation}
	\hat{\Psi}\,(t,x) = e^{\,i\alpha \,\hat{\Phi} (t,x)}\, \hat{\Psi}_{C}\,(t,x) \;.
\end{equation}

Hence, we have provided an example for which, guessing or determining the form of $\phi_{ml}$ allows the explicit computation of the statistical vector potential, the statistical factor and the duality transformation. In fact, the field theoretical formulation allows us to interpret the Jordan-Wigner string as a kink.\\

\paragraph{Two spatial dimensions.}
In two spatial dimension we shall recover the physics of conventional flux attachment and statistical transmutation as in the Abelian fractions of FQHE. This is, recovering the physics of the many-body Aharonov-Bohm effect. The natural angular variable is the polar angle, which can take values in $S^{1}$ and has a singularity at $\mathbf{x} = 0$, so the relevant homotopy group is $\pi_{1}(S^{1}) = \mathbb{Z}$. This can be formalised by means of the argument function  $\phi_{ml} \equiv \arg\,(\mathbf{\tilde{x}}_{ab}) = \arg\,(\mathbf{x}_{a} - \mathbf{x}_{b}\,; \hat{\text{e}}_{\mathbf{x}})$, where the angle is taken with respect to some arbitrary reference, in this case, the $x$-axis. The exchange property for this function reads $\phi_{ab} = \pm \pi + \phi_{ba}\,$. We can now compute the gauge potential as
\begin{flalign}\label{eq:2d_gaugefield}
\bm{a} \,(\mathbf{x}_{i}) &= \alpha\,\nabla_{\mathbf{x}_{i}}  \Phi\,(\mathbf{x}_{1},\dots,\mathbf{x}_{i},\dots,\mathbf{x}_{N}) = \alpha \, \bm{\nabla}_{\mathbf{x}_{i}} \bigg(\sum_{a<b} \phi_{ab}\bigg) = \alpha \, \bm{\nabla}_{\mathbf{x}_{i}} \Big[\sum_{i<b} \phi_{ib} + \sum_{a<i} \phi_{ai} \Big] \\
&= \alpha \, \bm{\nabla}_{\mathbf{x}_{i}} \Big[\sum_{i<b} \phi_{ib} + \sum_{a<i} \big(\pm \pi + \phi_{ia}\big) \Big] = \alpha \, \bm{\nabla}_{\mathbf{x}_{i}}  \Big[\sum_{j\neq i} \phi_{ij} \pm \sum_{a<i} \pi \Big] \\
&= \alpha\, \sum_{j\neq i} \bm{\nabla}_{\mathbf{x}_{i}} \arg\,(\mathbf{x}_{i} -\mathbf{x}_{j}\,; \hat{\text{e}}_{\mathbf{x}}) = \alpha\, \sum_{j\neq i} \bm{\nabla}_{\mathbf{x}_{i}} \arg\,(\tilde{\mathbf{x}}_{ij})\;, 
\end{flalign}
for which one can define a magnetic field and verify that there exist a relation with the charge density $n\,(\mathbf{x}_{i})$ of the form
\begin{equation} \label{eq:local_flux_attach}
	b\,(\mathbf{x}_{i}) = \bm{\nabla}_{\mathbf{x}_{i}} \times \bm{a} \,(\mathbf{x}_{i}) = \alpha \sum_{j\neq i} 2\pi\,\delta^{\,(2)} \,(\mathbf{x}_{i} - \mathbf{x}_{j}) \equiv 2\pi \alpha\, n\,(\mathbf{x}_{i})\,.
\end{equation}
In other words, we naturally recover local flux attachment. The associated gauge transformation in this case is commonly known in literature as a statistical or singular transformation, and reduces to 
\begin{equation}
	\Psi\,(\mathbf{x}_{1},\dots,\mathbf{x}_{N}) = e^{\,i\alpha \,\sum_{m<l} \text{arg} (\mathbf{x}_{m} - \mathbf{x}_{l}\,; \,\hat{\text{e}}_{\mathbf{x}})}\, \Psi_{c}\,(\mathbf{x}_{1},\dots,\mathbf{x}_{N}) \;,
\end{equation}
where the sum in the exponent is over all particles. Hence, for a given pairwise exchange of particles $i \leftrightarrow j\,$, where $1\le i < j \le N$, a corresponding $\pi$ phase from the argument function is collected by the composite wavefunction for every $w_{ab}^{-1}w_{ba}$ term $i\le a<b\le j$. This yields a statistical factor $\gamma_{ij} = \mp \alpha \pi \eta$, where $\eta \in \mathbb{Z}$ is the number of $a \leftrightarrow b$ possible pairs. This is nothing but a many-particle Aharonov-Bohm phase for flux $\alpha$.\\

Alternatively, such a peculiar choice of gauge potential is provided by construction, if the correct term is incorporated at the level of a field theoretical Lagrangian. We make use of the quantised Abelian Chern-Simons term at level $1/\alpha$ with $\alpha \in \mathbb{Z}$, minimally coupled to matter via source term
	\begin{equation}
		S = \frac{1}{4\pi \alpha} \int dt\,d^{2}\mathbf{x}\; \epsilon^{\,\mu \nu \lambda} \hat{a}_{\mu}\partial_{\nu} \hat{a}_{\lambda} - \int dt\,d^{2}\mathbf{x}\; \hat{J}^{\mu} \hat{a}_{\mu}\;.
	\end{equation}
	Computing the Euler-Lagrange equations for the gauge field in the presence of the matter source we are left with
	\begin{equation}
		\hat{J}^{\mu} = \frac{1}{2\pi \alpha} \,\epsilon^{\,\mu \nu \lambda} \partial_{\nu} \hat{a}_{\lambda}\;, \;\;\;\;\;\;\;\;\;\; \partial_{\mu}\, \hat{J}^{\mu} = 0\;,
	\end{equation}
	where the current time component becomes nothing but a constraint equation or Gauss's law of the form
	\begin{equation}\label{eq:gauss_law_app}
		\bm{\nabla} \times \hat{\bm{a}}\,(t,\mathbf{x}) = 2\pi \alpha \, \hat{n}\,(t,\mathbf{x})
	\end{equation}
	which we can attempt to solve in the Coulomb gauge $\bm{\nabla} \cdot \bm{a} = 0\,$. This allows us to write the vector potential as  $\bm{a} = \bm{\nabla} \times \varphi \,$, so that the Gauss's law becomes 
	\begin{equation}
		n\,(t,\mathbf{x}) = \frac{1}{2\pi \alpha} \bm{\nabla}^{2} \varphi\,(t,\mathbf{x})
	\end{equation}
	that can be solved using conventional Green's function methods to find that $\hat{\bm{a}}\,(t,\mathbf{x}) = \alpha \,\bm{\nabla} \,\hat{\Phi}\,(t,\mathbf{x}) $ and
	\begin{equation}
		\hat{\Phi}\,(t,\mathbf{x}) = \int d^{2}\mathbf{x}'\;\varphi\,(t,\mathbf{x} - \mathbf{x}')\,\hat{n}\,(t,\mathbf{x}')\;,
	\end{equation}
	where $\varphi \,(t,\mathbf{x}) = \tan ^{-1}\,(y/x)$ is the conventional polar angle, and where we have assumed that the ``charge'' density is point-like so that the Chern-Simons gauge potential can be written as a pure gauge. This is nothing but the field theoretical version of flux attachment previously found in first-quantised language. The corresponding singular gauge transformation is
 \begin{equation}
	\hat{\Psi}\,(t,\mathbf{x}) = e^{\,i\alpha \,\hat{\Phi} (t,\mathbf{x})}\, \hat{\Psi}_{C}\,(t,\mathbf{x}) \;.
\end{equation}\\

\paragraph{Three spatial dimensions.}	
We shall generalise the previous prescription in the group of three-dimensional rotations, parametrised now by two angles or an axis-angle complex. Here we are looking for an analogue of the argument function, so the natural choice to make is the solid angle $\Omega\,(\mathbf{x}_{m} - \mathbf{x}_{l})$. This function will acquire values in $[0,4\pi)$ and will have a Dirac string singularity at $\mathbf{x} = \mathbf{x}_{0}\,$, which cannot be removed via gauge transformation. The appropriate homotopy group is non-trivially $\pi_{1}(\text{SO}(3)) = \mathbb{Z}_{2}$. This is slightly more convoluted than previous cases as not only do we need a reference axis $\hat{\text{e}}_{\mathbf{n}}$ with reference to $\mathbf{x}_{m}$ in order to define an angle, but also a reference closed contour $\Gamma$ around which $\mathbf{x}_{l}$ is transported in order to define a 3d section (or slice), and the relative orientation between references. A simple choice can be $\hat{\text{e}}_{z}$, $\Gamma$ to lie in the $x-y$ plane (i.e. perpendicular to the axis), and for it to be a circle, so that the resulting solid angle measures the area subtended by a conical section. Despite this discussion, we have not been able to construct this object in closed and \textit{local} form, but we point to Refs. \cite{marino2017quantum,kleinert2008multivalued,haldane1986ferro} as potential sources of inspiration for the interested reader.

\section{Making Sense of the Statistical Flux Operator} \label{sec:sense}
\begin{figure}[h]
\centering
    \includegraphics[width=\textwidth]{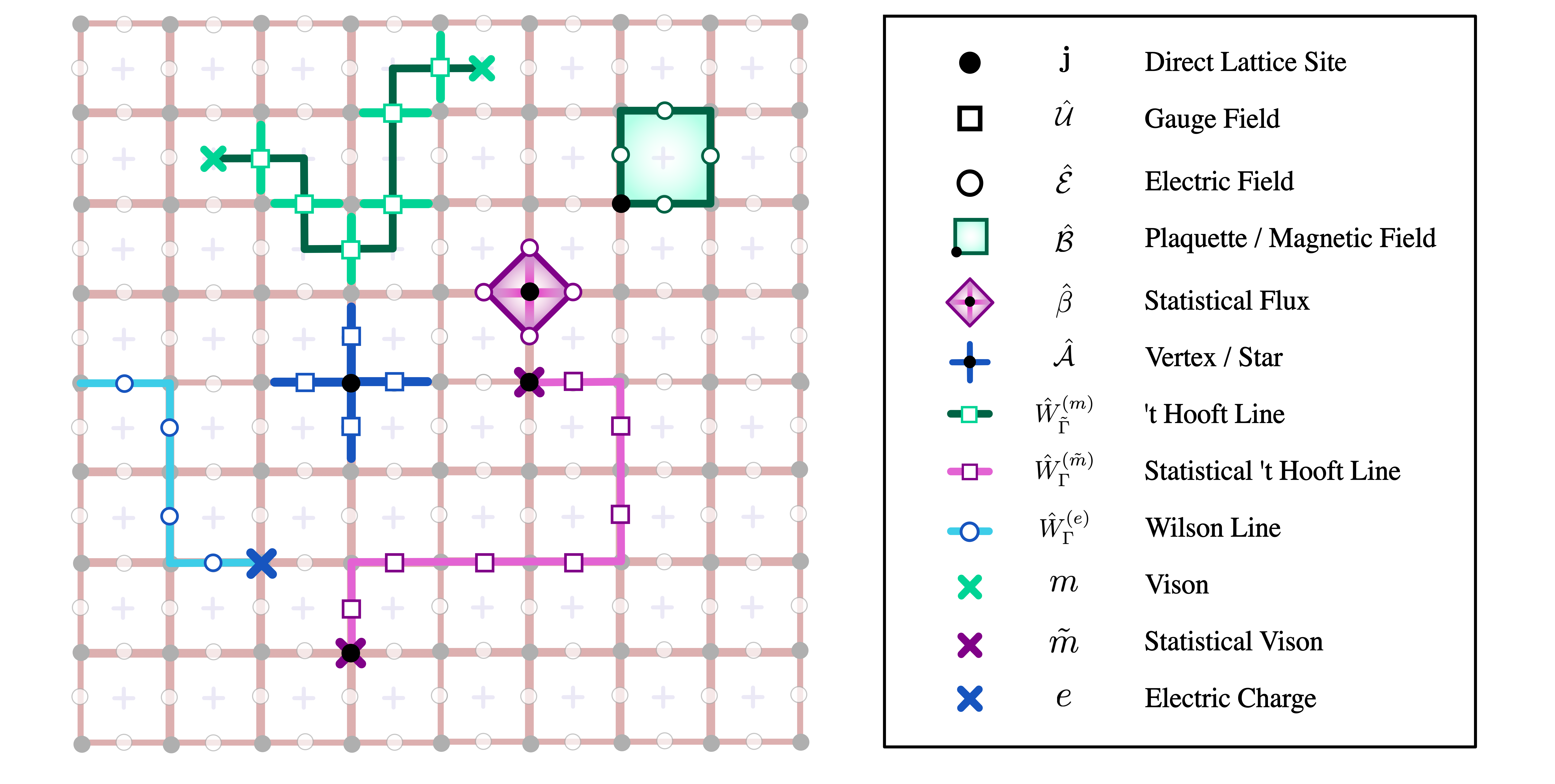}
    \caption{Gauge-theoretic operators on a square lattice. Observe the statistical operators as opposed to the more conventional electromagnetic operators.}
    \label{fig:plaquette_stuff}
\end{figure}

It is worth comparing morphology and effect of the statistical flux operator with respect to more familiar electromagnetic operators on the lattice. Let us discuss a particular case for a square lattice in $d=2$ and a $\mathbb{Z}_{2}$ gauge group. We define the link variable to be $\hat{\mathcal{U}}_{\mu}(\mathbf{j}) \equiv \hat{\sigma}^{x}_{\mu}(\mathbf{j})$. In this context, abusing language, one might refer to link variables as gauge fields. In addition, electric fields are given by some other link variables, namely $\hat{\mathcal{E}}_{\mu}(\mathbf{j}) \equiv \hat{\sigma}^{z}_{\mu}(\mathbf{j})$. Finally, magnetic fields or fluxes are given by the plaquette operator
\begin{equation}
    \hat{\mathcal{B}}_{\,\tilde{\mathbf{j}}} = \prod_{\braket{\tilde{\mu}}} \hat{\mathcal{U}}_{\tilde{\mu}}(\tilde{\mathbf{j}}) \equiv \prod_{\mathbf{j} \, \in \, \square} \hat{\sigma}^{x}_{\mu}(\mathbf{j})\;,
\end{equation}
which is the product of dual links for a given dual lattice site, or equivalently, the product of gauge fields over a real-space plaquette. We can also define a \textit{star} or \textit{vertex operator} accounting for the electric flux in a similar manner
\begin{equation}
    \hat{\mathcal{A}}_{\,\mathbf{j}} = \prod_{\braket{\mu}} \hat{\mathcal{E}}_{\mu}(\mathbf{j}) = \prod_{\mathbf{j}\,\in\, \boldsymbol{+}} \hat{\sigma}^{z}_{\mu}(\mathbf{j})\;.
\end{equation}
In other words, electric fluxes live on the real lattice while magnetic fluxes live on the dual lattice. See Figures \ref{fig:lattice_dual} and \ref{fig:plaquette_stuff} for a graphical representation of the notation so far. Then, quantum double models such as the Toric Code are simply written as ``electromagnetism in the absence of matter", meaning
\begin{equation}
    \hat{H} = - \sum_{\mathbf{j}} \hat{\mathcal{A}}_{\,\mathbf{j}} - \sum_{\tilde{\mathbf{j}}} \hat{\mathcal{B}}_{\,\tilde{\mathbf{j}}} = - \sum_{\mathbf{j}} \bigg( \prod_{\mathbf{j}\,\in\, \boldsymbol{+}} \hat{\sigma}^{z}_{\mu}(\mathbf{j}) + \prod_{\mathbf{j}\, \in \, \square} \hat{\sigma}^{x}_{\mu}(\mathbf{j}) \bigg) \;.
\end{equation}
We now clearly see the difference with the statistical flux operator as it is
\begin{equation}
    \hat{\beta}_{\,\mathbf{j}} \equiv \prod_{\braket{\mu}} \hat{\mathcal{U}}_{\mu}(\mathbf{j}) =\prod_{\mathbf{j} \, \in \,\boldsymbol{+}} \hat{\sigma}^{x}_{\mu}(\mathbf{j})\;
\end{equation}
in this context. It is the dual magnetic field or the twisted (or magnetic) analogue of the electric vertex operator. It counts the magnetic flux in the direct lattice sites and yet, it is different to the conventional magnetic flux operator. In the presence of matter on the sites then, we see that Eq. \eqref{eq:gauss_flux} is nothing but the magnetic Gauss' law counting monopoles. In other words, it is lattice flux attachment or the incarnation of Chern-Simons gauge theory for a discrete system and local $\mathbb{Z}_{2}$ symmetry. Notice that the statistical flux operator is gauge invariant in the same way the conventional magnetic flux is.\\

\paragraph*{Excitations.}
Furthermore, we can study excitations for such a system. Before launching into the discussion it is worth stressing that there is no matter in this example, so all the possible excitations are ``electromagnetic'' in nature. In order to find those, we must define the electric and magnetic string operators, also known as Wilson lines
\begin{equation}
\hat{W}^{(e)}_{\Gamma} = \prod_{\mu \,\in \, \Gamma} \hat{\sigma}^{x}_{\mu}
\end{equation}
and 't Hooft lines
\begin{equation}
\hat{W}^{(m)}_{\tilde{\Gamma}} = \prod_{\mu \,\in \, \tilde{\Gamma}} \hat{\sigma}^{z}_{\mu}
\end{equation}
respectively. In a given vacuum or ordered background, if lines are open, a pair of electric $e$ (magnetic $m$) excitations are found at the ends of the Wilson ('t Hooft) strings. Hence a disorder operator, namely a $\mathbb{Z}_{2}$ magnetic vortex-antivortex linear combination or \textit{vison} can be defined as 
\begin{equation}
    \hat{\tilde{\mathcal{V}}}^{\dagger}(\tilde{\mathbf{j}}) \equiv \hat{\tau}^{x}(\tilde{\mathbf{j}})= \prod_{(\tilde{\mu},\tilde{\mathbf{k}}) \,\in\, \tilde{\Gamma} < \tilde{\mathbf{j}}} \hat{\sigma}^{z}_{\tilde{\mu}} (\tilde{\mathbf{k}})\;,
\end{equation}
which defines the excitation $m$. This works as a dual Jordan-Wigner string extending from the edge of the system up to $\tilde{\mathbf{j}}$ and following some path $\tilde{\Gamma}$. We note that an open string of the form
\begin{equation}
      \prod_{ \tilde{\mathbf{i}}\le(\tilde{\mu},\tilde{\mathbf{k}}) \,\in\, \tilde{\Gamma} < \tilde{\mathbf{j}}} \hat{\sigma}^{z}_{\tilde{\mu}} (\tilde{\mathbf{k}})\;.
\end{equation}
will feature two visons at its endpoints, located at $\tilde{\mathbf{i}}$ and $\tilde{\mathbf{j}}\,$ (see Figure \ref{fig:plaquette_stuff}). Analogously, the associated 't Hooft line for the excitations of the statistical flux operator is
\begin{equation}
\hat{W}^{(\tilde{m})}_{\Gamma} = \prod_{\mu \,\in \, \Gamma} \hat{\sigma}^{z}_{\mu}\;,
\end{equation}
in terms of which we can define a twisted or statistical vison 
\begin{equation}
    \hat{\mathcal{W}}^{\dagger}(\mathbf{j}) \equiv \prod_{(\mu,\mathbf{k}) \,\in\, \Gamma < \mathbf{j}} \hat{\sigma}^{z}_{\mu} (\mathbf{k})\;,
\end{equation}
which defines the excitation $\tilde{m}$. This is nothing but our conventional Jordan-Wigner string, which we observe to be a ``gauge string''. It is worth pursuing this example a little further and realise that in the presence of matter fields living in the direct lattice sites $\hat{c}_{\mathbf{j}}$, a configuration such as 
\begin{equation}
    \hat{c}_{\mathbf{i}}^{\dagger} \;\Bigg( \,\prod_{ \mathbf{i}\le ( \mu,\mathbf{k}) \,\in\, \Gamma < \mathbf{j}} \hat{\sigma}^{z}_{\mu} (\mathbf{k}) \Bigg)\; \hat{c}_{\mathbf{j}}
\end{equation}
describes a pair of composite particles formed by matter attached to, respectively $\mathbb{Z}_{2}$ statistical gauge vortices and antivortices, and located at sites $\mathbf{i}$ and $\mathbf{j}$. These composites interact non-locally through the gauge string $\Gamma$. The string tension causes an energy penalty and, in the usual sense, allows these composites to be either confined or free. We see they are, actually, the lattice $\mathbb{Z}_{2}$ closest relatives to Zhang-Hansson-Kivelson (ZHK) vortices \cite{zhang1989effective,zhang1995chern,horvathy2009vortices}. This is flux-attachment considerations at play or, more generally, composite-particle-duality ones. It is worth noting that these configurations are excitations of the system are similar to the so-called $e-m$ pairs or ``dyons'' from the Toric Code, which are composite fermions according to the fusion rule $ e\;\times\;m = f\,.$ These are related in spirit to the composites just discussed but, as it is palpable, they are different objects. However, we could envision, using a similar language, some new $e-\tilde{m}$ composites that could be \textit{emergent dyons} on a lattice setup, in reference and as an extension of the concept introduced in Section \ref{sec:genFluxsec}. \\

Then, what is the relation of quantum double models like the Toric Code and the models we propose? The toric code is a \textit{twisted doubled} version of the type of models we introduce. Schematically, this can be seen provided that the Toric Code is nothing but the lattice pure Chern-Simons limit of a doubled or mutual $\text{U}(1)\times \text{U}(1)$ Maxwell-Chern-Simons theory coupled to matter \cite{kou2008mutual,olesen2015doubled}. When the matter field condenses, it enters a $\mathbb{Z}_{2}$ spin liquid phase and the $\text{U}(1)$ symmetry breaks into $\mathbb{Z}_{2}$ \cite{freedman2004class}. This corresponds to the pure doubled-Chern-Simons term
\begin{equation}
    S_{\text{2CS}} = \frac{1}{4\pi} \int d^{3}x\;K_{IJ}\,\epsilon^{\mu \nu \lambda} a^{I}_{\mu} \partial_{\nu} a^{J}_{\lambda}
\end{equation}
for an off-diagonal K-matrix $K_{IJ} \equiv \hat{\mathbf{K}} = 2 \hat{\sigma}^{x}\,$. The presence of two mutual Chern-Simons dynamics translates into $e$ and $m$ being mutual \textit{semions}. Our model, on the contrary, there are no mutual semions, but there are anyonic composites. The theory is trivial in the absence of matter, and it is the lattice limit of a single conventional Chern-Simons term. In a sense, a ``twisted square-root'' of the Toric Code. Yet, the generation of composites in both our model and the Toric Code follow an order-disorder structure. A final comment is that the discussion here can be extended in all generality to different symmetry groups $G$, lattices and dimensionalities. This, once again will correspond to different avatars of flux-attached gauge theories.


\section{Hidden Gauge Theories in Quantum Spin Models}
\label{sec:hidden}

\paragraph{On the quantum Heisenberg Chain.} Let us illustrate that the composite dual of the spin--$\frac{1}{2}$ quantum Heisenberg model in $d=1$ is, secretly, a lattice gauge theory. We consider the Heisenberg Hamiltonian with anisotropy or transverse field $\gamma$ with
\begin{equation}
        \hat{H} = J \sum_{j=1}^{N} \Big( \hat{S}^{x}_{j} \hat{S}^{x}_{j+1} + \hat{S}^{y}_{j} \hat{S}^{y}_{j+1} \Big) + \gamma J \sum_{j=1}^{N} \hat{S}^{z}_{j} \hat{S}^{z}_{j+1}\;.
\end{equation}
Defining $\hat{S}^{\pm} = \hat{S}^{x} \pm i \hat{S}^{y}$ as the spin ladder operators for $\hat{S}^{a} = \frac{1}{2} \hat{\sigma}^{a}$, we find relations
\begin{flalign}
 & \hat{S}^{+} \hat{S}^{-} = \frac{1}{2} + \hat{S}^{z} \;,\\
 & \hat{S}^{-} \hat{S}^{+} = \frac{1}{2} - \hat{S}^{z}
\end{flalign}
We can now re-express the Hamiltonian as
\begin{equation}
        \hat{H} = \frac{J}{2} \sum_{j=1}^{N} \Big( \hat{S}^{+}_{j} \hat{S}^{-}_{j+1} + \hat{S}^{-}_{j} \hat{S}^{+}_{j+1} \Big) + \gamma J \sum_{j=1}^{N} \Big( \hat{S}^{+}_{j}\hat{S}^{-}_{j} - \frac{1}{2} \Big) \, \Big( \hat{S}^{+}_{j+1}\hat{S}^{-}_{j+1} - \frac{1}{2} \Big)\;.
\end{equation}
We can now define our bare operators $\hat{c}_{j}\equiv \hat{S}^{-}_{j}$ and $\hat{c}^{\dagger}_{j} \equiv \hat{S}^{+}_{j}$. The disorder operator is a kink operator and the Jordan-Wigner brane is the usual string. In $d=1$ the 't Hooft line and JW string are essentially the same object. Explicitly, 
\begin{equation}
    \hat{\mathcal{W}}_{\tilde{j}}\,(\Gamma) = \exp{\big[i\,\hat{\Phi}_{\tilde{j}}\,(\Gamma)\big]} =  \exp{\Big(i\alpha \sum_{l=1}^{N} \Theta (j-l) \, \hat{n}_{l} \Big)} = \exp{\Big(i\alpha \sum_{l=1}^{j-1} \hat{n}_{l} \Big)}\;,
\end{equation}
where the links or dual lattice sites are encapsulated in $\tilde{j} \equiv (j,\mu) \equiv (j-l)$ in the lattice Heaviside step function $\Theta$. We thus see that the hopping term transforms according to
\begin{equation}\label{eq:hoping}
    \hat{S}^{+}_{j} \hat{S}^{-}_{j+1} = \hat{c}^{\dagger}_{j} \hat{c}_{j+1} =  \hat{f}^{\dagger}_{j} \,\hat{\mathcal{W}}^{\dagger}_{\tilde{j}} \,\hat{\mathcal{W}}_{\tilde{j}+1}\, \hat{f}_{j+1} =  \hat{f}^{\dagger}_{j} e^{i\alpha \big(\sum_{l=1}^{j-1}\hat{n}_{l} - \sum_{m=1}^{j}\hat{n}_{m} \big)} \hat{f}_{j+1} = \hat{f}^{\dagger}_{j} e^{-i\alpha \hat{n}_{j}} \hat{f}_{j+1}\;.
\end{equation}
This defines an operator-valued Peierls phase
\begin{equation}
\hat{a}_{x}(j) = \hat{a}\,(j, j+\hat{\text{e}}_{x}) = \frac{1}{\hbar} \int_{j}^{j+1} dx\;\hat{a}_{x}(x) = - \alpha \hat{n}_{j}\;,
\end{equation}
which is nothing but the lattice statistical gauge field satisfying the Gauss's law
\begin{equation}\label{eq:gauss_flux_heis}
    \hat{G}_{j} \ket{\text{Phys}} \equiv  \hat{a}_{x}(j) \,\ket{\text{Phys}} = - \alpha \hat{n}_{j} \,\ket{\text{Phys}} = 0\;.
\end{equation}
Hence, the composite dual model to the Heisenberg chain is the constrained Hamiltonian
\begin{flalign}
\begin{cases}
            &\hat{H} = \frac{J}{2} \sum_{j=1}^{N} \big( \hat{f}^{\dagger}_{j} e^{i \hat{a}_{x}(j)}\hat{f}_{j+1} + \hat{f}_{j} \,e^{ - i \hat{a}_{x}(j)} \hat{f}^{\dagger}_{j+1} \big) + \gamma J \sum_{j=1}^{N} \big( \hat{f}^{\dagger}_{j}\hat{f}_{j} - \frac{1}{2} \big) \, \big( \hat{f}^{\dagger}_{j+1}\hat{f}_{j+1} - \frac{1}{2} \big)\;,\\
            \\
            &\big[ \hat{a}_{x}(j) + \alpha \hat{n}_{j} \big] \,\ket{\text{Phys}} = 0\;.
\end{cases}
\end{flalign}
This is a lattice gauge theory of interacting composite matter fields $\hat{f}$ coupled to a statistical gauge potential. They are ``gauge-dressed'' by the Jordan-Wigner string. It is worth noting that $\hat{f}$ fields are \textit{anyonic} in nature, meaning that their statistics is governed by $\alpha$, and have a hard core since they anticommute on the same site. Hence, the Heisenberg chain is equivalent to a theory of linear Abelian anyons in the presence of effective dynamical electric field $\sim \partial_{t}\hat{n}_{j}$. As it turns out, this model has a particularly simple expression for $\alpha = \pi$, since the composite fields can be identified with hard-core spinless fermions, meaning that they satisfy 
\begin{flalign}
    & \{ \hat{f}_{i}, \hat{f}_{j}\} = \{ \hat{f}^{\dagger}_{i}, \hat{f}^{\dagger}_{j}\} = 0\;, \\
    & \{ \hat{f}_{i}, \hat{f}^{\dagger}_{j} \} = \delta_{ij}\;.
\end{flalign}
Then, the operator-valued Peierls phase can be expressed in a series expansion
\begin{equation}
    e^{i\hat{a}_{x}(j)} = e^{-i\pi \hat{f}^{\dagger}_{j} \hat{f}_{j}} = 1- 2\,\hat{f}^{\dagger}_{j} \hat{f}_{j} 
\end{equation}
so that the original hopping \eqref{eq:hoping} now becomes
\begin{equation}
    \hat{S}^{+}_{j} \hat{S}^{-}_{j+1} = \hat{f}^{\dagger}_{j} e^{-i\pi \hat{n}_{j}} \hat{f}_{j+1} = \hat{f}^{\dagger}_{j} (1- 2\,\hat{f}^{\dagger}_{j} \hat{f}_{j} ) \hat{f}_{j+1} = \hat{f}^{\dagger}_{j} \hat{f}_{j+1} \;.
\end{equation}
In this case, the Peierls phase has no effect. Hence, the original Heisenberg model is mapped to an interacting theory of spinless fermions
\begin{equation}
        \hat{H} = \frac{J}{2} \sum_{j=1}^{N} \big( \hat{f}^{\dagger}_{j} \hat{f}_{j+1} + \text{H.c.} \big) + \gamma J \sum_{j=1}^{N} \big( \hat{n}_{j} - \frac{1}{2} \big) \, \big( \hat{n}_{j+1} - \frac{1}{2} \big) \;\;\;\;\;\;+\;\; \text{Boundary Term},
\end{equation}
it is \textit{fermionised} and can be solved via Bethe ansatz. In the absence of anisotropy ($\gamma=0$), the system reduces to the quantum 1d XY model and fermions become free. We can then diagonalise the Hamiltonian in momentum space as usual. 
\vspace{0.3cm}
\noindent
\paragraph{On the quantum Ising model.} The above reasoning has also implications for the classical 2d Ising model, which is equivalent to the 1d quantum (or transverse) Ising model through transfer matrix formalism \cite{fradkin2013,shankar2017quantum,sachdev1999quantum}. The corresponding Hamiltonian is
\begin{equation}
    \hat{H} = - J \sum_{j} \hat{\sigma}^{z}_{j} \hat{\sigma}^{z}_{j+1} - J\gamma \sum_{j} \hat{\sigma}^{x}_{j}\;.
\end{equation}
This model is electric-magnetic lattice self-dual $\hat{\sigma}_{j} \leftrightarrow \hat{\mathcal{W}}_{\,\tilde{j}}\;$. Here, it is convenient\footnote{This is because $\hat{\sigma}^{x} = \hat{c}^{\dagger} + \hat{c}\;$, which is highly non-local. Alternatively, we can consider a non-rotated Jordan-Wigner transformation but a rotated Ising model $\hat{H}' = - J\sum_{j}\hat{\sigma}^{z}_{j} - J\gamma \sum_{j}\hat{\sigma}^{x}_{j} \hat{\sigma}^{x}_{j+1} \,$ \cite{kogut1979introduction}.} to perform a $\pi/2$ spin rotation or global transformation around the $y$-axis, which yields $\hat{\sigma}^{z}\rightarrow \hat{\sigma}^{x}$ and $\hat{\sigma}^{x} \rightarrow -\hat{\sigma}^{z}$. So the raising and lowering operators are redefined as
\begin{equation}
    \hat{\sigma}^{\pm} = -\frac{1}{2} (\hat{\sigma}^{z} \mp i\hat{\sigma}^{y})
\end{equation}
The rotated mapping now yields a non-local operator $\hat{\sigma}^{z}_{j} = - (\hat{\sigma}^{+}_{j} + \hat{\sigma}^{-}_{j})\equiv - (\hat{c}^{\dagger}_{j} + \hat{c}_{j} )\;$. Writing in the composite basis gives
\begin{flalign}
    &\hat{\sigma}^{x}_{j} = 1 - 2\,\hat{n}_{j}\;,\\
    &\hat{\sigma}^{z}_{j} = (\hat{f}^{\dagger}_{j} + \hat{f}_{j})\,\hat{\mathcal{W}}_{\tilde{j}}\,(\Gamma)\;.
\end{flalign}
The quantum Ising model can be re-expressed in the composite dual anyonic picture, yielding again a lattice gauge theory, albeit a non-local one due to the non-conservation of anyon number. This is, non-conserving anyon terms give rise to JW strings, which we identify as \textit{'t Hooft} lines that extend to infinity or the edge of the system. When these anyons coincide with spinless hard-core fermions, the system is local and quadratic in (composite) fermion fields, namely
\begin{equation}
    \hat{H} = - J \sum_{j} \Big[\hat{f}^{\dagger}_{j}\hat{f}_{j+1} + \hat{f}^{\dagger}_{j+1}\hat{f}_{j} + \hat{f}^{\dagger}_{j}\hat{f}^{\dagger}_{j+1} + \hat{f}_{j+1}\hat{f}_{j} - 2\gamma\,\Big(\hat{n}_{j} -\frac{1}{2}\Big)\Big]  \;\;\;\;\;\;+\;\; \text{Boundary Term},
\end{equation}
but with anomalous BCS-pairing-like terms. In fact, the composite dual becomes an instance of the Kitaev p-wave topological superconducting chain, written in more generality as
\begin{equation}
    \hat{H}_{\text{Kitaev}} =  \sum_{j} \Big[ -t\,\Big(\hat{f}^{\dagger}_{j}\hat{f}_{j+1} + \hat{f}^{\dagger}_{j+1}\hat{f}_{j} \Big) + \Delta \,\hat{f}^{\dagger}_{j}\hat{f}^{\dagger}_{j+1} + \Delta^{*}\,\hat{f}_{j+1}\hat{f}_{j} - \mu \, \Big(\hat{n}_{j} -\frac{1}{2}\Big)\Big] \;,
\end{equation}
for $t=-\Delta$ and $\Delta \in \mathbb{R}$. This model, in turn, can be re-expressed in terms of Majorana (real) operators. Hence, either in terms of Majorana or conventional Dirac (complex) fermion operators, the model can be diagonalised in momentum space by means of conventional Bogoliubov transformation. Hence, the resulting quasiparticle operators will be \textit{composite bogoliubons}, namely Majoranas. Study of the energy spectrum indicates a topological (i.e. gap-closing) phase transition. There is a well-known correspondence between topological/trivial phases in the composite dual model and the ordered/disorder symmetry-breaking phases in the original Ising model. This is not by chance, but because we have performed the 1+1D equivalent to flux attachment by means of the Jordan-Wigner transformation. It is worth noting that the system does not preserve the number of composite fermions $[\hat{N}, \hat{H}] \neq 0$ due to the anomalous terms, but it preserves fermionic parity operator $\hat{P} \equiv (-1)^{\hat{N}} = \exp{\big(i\pi \sum_{j}\hat{n}_{j}\big)}$, meaning $[\hat{P},\hat{H}] = 0$ since fermion operators come in pairs. The number of Majoranas is not conserved but the fact that they come in pairs is, i.e. they are the excitations at the edges of strings. In our current view, the conservation of fermion parity is a global by-product of gauge invariance, since $\hat{P} = \prod_{j}\hat{G}_{j}\,$, something which has not been acknowledged in literature.  Careful attention to boundary conditions \cite{fradkin2013} indicates that for each theory of spins there is a fermionic parity $\mathbb{Z}_{2}$--ambiguity on the boundary conditions for the composite counterpart, meaning the \textit{composite} $\longrightarrow$ \textit{bare} mapping is 2 $\longrightarrow$ 1. In fact, this shows up in the form of spin domain wall pairs in the Ising model, or as a pair of Majorana zero-modes at the edges in the composite picture. Now, it is worth thinking about this model in its classical 2d version, as the presence of defects can be seen as a classical $\mathbb{Z}_{2}$ gauge theory, as pointed out by Kadanoff and Ceva \cite{kadanoffceva1971}. In this case, Onsager's ``classical composite fermions'' are nothing but a consequence of classical flux attachment. This 2-to-1 ambiguity appears there provided the Onsager fermion is an order-disorder composite. Hence two links --- a.k.a. two dual sites --- and, in turn, two corresponding disorder variables living there --- a.k.a. magnetic vortices ---, can be attached to each spin matter field living on the sites. This $\mathbb{Z}_{2}$ local ambiguity is a $\mathbb{Z}_{2}$ gauge invariance. All the above, has a $\mathbb{Z}_{N}$ counterpart \cite{fradkin2017disorder}, where the Onsager fermions now become \textit{parafermions}, and the Majorana zero-modes are now parafermion zero-modes.\\

Similar remarks apply to the classical Ising lattice gauge theory in 3d, which is equivalent to a quantum Ising (or $\mathbb{Z}_{2}$) lattice gauge theory in 2d, and electric-magnetic lattice dual $\hat{\sigma}_{\mathbf{j}} \leftrightarrow \hat{\mathcal{W}}_{\,\tilde{\mathbf{j}}}\,$ to a 2d quantum Ising model. In the quantum $\mathbb{Z}_{2}$ LGT, visons appear at the ends of an open 't Hooft line. Composite fields there are the result of attaching of such visons to matter, namely ``charges'' or spin fields residing on the sites, which provides a quantum version of Onsager's composite fermions.

\bibliography{references}
\bibliographystyle{apsrev4-2}

\end{document}